\documentclass[aps,superscriptaddress,twocolumn]{revtex4}

\usepackage[pdftex]{graphicx}
\usepackage{dcolumn}
\usepackage{bm}
\usepackage{color}
\usepackage{amsmath}
\usepackage{nicefrac}
\usepackage{amssymb}

\newcommand{\ff}[1]{{\boldsymbol #1}}
\newcommand{\ca}[1]{{\cal #1}}
\newcommand{\bi}{\begin{itemize}}
\newcommand{\ei}{\end{itemize}}
\newcommand{\be}{\begin{equation}}
\newcommand{\ee}{\end{equation}}
\newcommand{\ba}{\begin{eqnarray}}
\newcommand{\ea}{\end{eqnarray}}

\newcommand{\dm}{\bm{\hat{\rho}}}

\DeclareMathOperator{\tr}{tr}

\usepackage[pdftex,colorlinks=true,linkcolor=blue,citecolor=blue,filecolor=blue]{hyperref}

\begin{document} 
  
\title{Accessing long timescales in the relaxation dynamics of spins coupled to a conduction-electron system using absorbing boundary conditions}

\author{Michael Elbracht} 

\affiliation{I. Institute of Theoretical Physics, Department of Physics, University of Hamburg, Jungiusstra\ss{}e 9, 20355 Hamburg, Germany}

\author{Michael Potthoff}

\affiliation{I. Institute of Theoretical Physics, Department of Physics, University of Hamburg, Jungiusstra\ss{}e 9, 20355 Hamburg, Germany}

\affiliation{The Hamburg Centre for Ultrafast Imaging, Luruper Chaussee 149, 22761 Hamburg, Germany}

\begin{abstract}
The relaxation time of a classical spin interacting with a large conduction-electron system is computed for a weak magnetic field, which initially drives the spin out of equilibrium.
We trace the spin and the conduction-electron dynamics on a time scale, which exceeds the characteristic electronic scale that is set by the inverse nearest-neighbor hopping by more than five orders of magnitude.
This is achieved with a novel construction of absorbing boundary conditions, which employs a generalized Lindblad master-equation approach to couple the edge sites of the conduction-electron tight-binding model to an external bath. 
The failure of the standard Lindblad approach to absorbing boundaries is traced back to artificial excitations  initially generated due to the coupling to the bath. 
This can be cured by introducing Lindblad parameter matrices and by fixing those matrices to perfectly suppress initial-state artifacts as well as reflections of physical excitations propagating to the system boundaries. 
Numerical results are presented and discussed for generic one-dimensional models of the electronic structure.
\end{abstract} 

\maketitle 

\section{Introduction}
\label{sec:intro}

The relaxation of a nonequilibrium state of a single or several local magnetic moments is one of the central issues in various atomistic spin-dynamics theories \cite{TKS08,SHNE08,BMS09,FI11,EFC+14}.
In many cases the local moments are treated as classical spins and the relaxation process is covered by an atomistic version of the Landau-Lifshitz-Gilbert (LLG) equation \cite{llg}.
Such effective spin-only theories are extremely effective and have proven to be very successful.

In many cases, however, an explicit treatment of the coupling of the spins to the conduction-electron system is necessary and can be described, e.g., with $s$-$d$-type models \cite{VZ}. 
Those approaches comprise the effective spin-only theories and can rederive the LLG equation and the Gilbert-damping parameter using, e.g., perturbative techniques \cite{ON06,BNF12,UMS12,BN19}, or perturbative or other downfolding approaches within a first-principles framework \cite{AKvSH95,KK02,CG03,EMKK11,Sak12}.

An explicit and non-perturbative treatment of the full problem of coupled spin and electron dynamics on equal footing becomes necessary, if the exchange interaction $J$ between the spin and the conduction-electron system is strong, if the spins are driven fast compared to typical electronic time scales, or, generally speaking, if the coupled dynamics of spin and electron degrees of freedom is intricate and cannot be separated easily.
Examples comprise one-dimensional systems, where the perturbative derivation of Gilbert damping breaks down \cite{SP15}, or spin pre-relaxation effects due to electronic correlations \cite{SRP16a}, or the feedback of local topological properties of the fast electron system to the slow spin dynamics \cite{SP17,EMP20,BN20}.
Certainly, another general motivation to address the full problem is the discovery of new physical phenomena.

\begin{figure}[b]
\includegraphics[width=0.6\columnwidth]{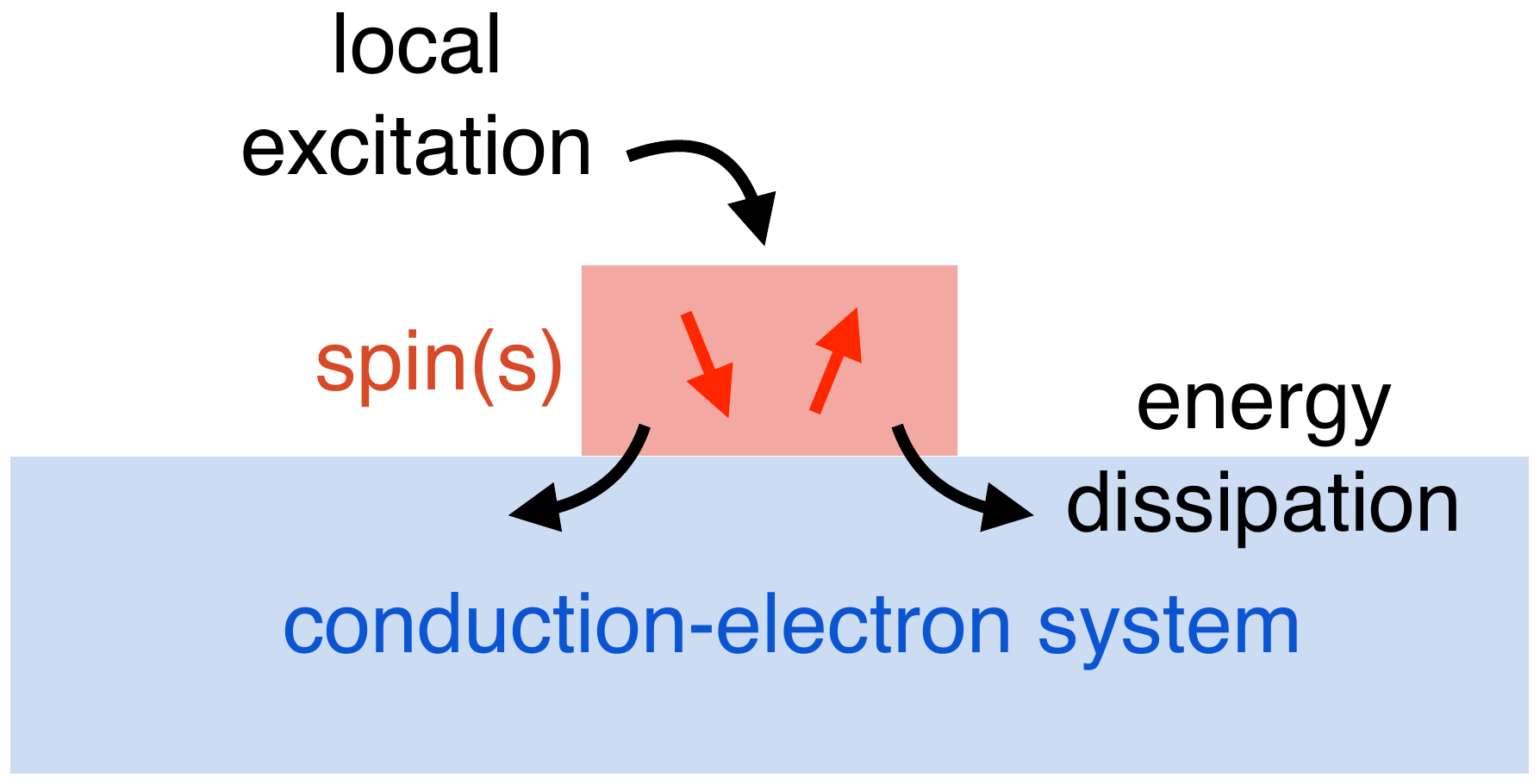}
\caption{
Relaxation of a single spin or a few spins interacting with a large conduction-electron system after an initial local excitation.
In the long-time limit, the spin-electron system is expected to reach its ground state locally, i.e., in the vicinity of the impurity spin(s), since the excitation energy is completely dissipated to the bulk.
}
\label{fig:relax}
\end{figure}

With the present work we would like to focus on the technical aspects and the numerical feasibility of a full, combined treatment of spin and electron degrees of freedom for a particular class of problems, as sketched in Fig.\ \ref{fig:relax}.
We consider a single classical spin (or a few spins) coupled to a finite but large system of noninteracting electrons described by a tight-binding model with nearest-neighbor hopping on a lattice of $L$ sites.
A one-dimensional geometry is assumed for simplicity but the discussion will be general.
The coupling is given by a local exchange interaction $J$ at a site $i_{0}$ of the lattice, and the system is assumed to be instantaneously kicked out of its ground state by some strong but local perturbation at the same site.
There is a closed system of equations of motion \cite{SP15} determining the real-time dynamics such that, in principle, this type of problem can be solved (numerically) exactly. 
One expects that locally the system decays to its ground state, i.e., that all local observables in the vicinity of $i_{0}$ converge to their ground-state values as time $t\mapsto \infty$.
For a single classical spin, the time scale required for the completion of this process defines the spin-relaxation time $\tau$.
Our goal is the numerically exact computation of $\tau$ and of other local observables in the interaction region close to $i_{0}$ by solving the equations of motion for coupled spin and electron dynamics explicitly.

While this type of calculation provides the maximum information on the system, it runs into computational troubles, when the relevant time scale, e.g., the spin relaxation time, becomes large compared to $L / v$, where $v$ is the characteristic velocity, at which energy- and spin-carrying excitations propagate through the electron system.
Namely, since energy and spin are conserved quantities, the excitation energy and the excess spin must be completely transported away from $i_{0}$ during the relaxation process and must be fully dissipated into the macroscopically large electron system.
Thus, the dissipation rate sets a bound on $\tau$.
As the computational effort scales about cubic with the system size $L$, long-time relaxation processes cannot be treated exactly. 

Calculations are spoiled by unwanted reflections of excitations, which back-propagate and interfere with the system dynamics in the interaction region.
This type of problem is well known in atomic, molecular and optical physics, where an unbound quantum system under study is conceptually decomposed into an interaction region of finite spatial extent and an asymptotic region where the (single-particle) wavefunction has some asymptotic form, and where it is desirable to focus on the dynamics in the interaction region only.
This can be achieved by imposing absorbing boundary conditions (absorbing BC), which minimize reflections from the edge of the core physical system represented on a numerical grid \cite{AAB+08}.
In most cases, one uses a complex absorbing potential (CAP) as an additional non-Hermitian term in the Hamiltonian, which is optimized with respect to its reflection properties \cite{Man02}. 
In the context of wave equations this is also known as perfectly matched layers \cite{Ber94}.
Such techniques are widely used but become problematic for systems with more than a single quantum particle \cite{SK10} since, if particles are lost, the Schr\"odinger equation with a CAP is not able to consistently describe the remainder of the system.

A consistent formalism can be based on Markovian quantum master equations of the Lindblad type \cite{Lin76,Pea12}, which focus on the many-body statistical operator $\dm(t)$ rather than on the single-particle wavefunction of the quantum system and which preserve the trace, Hermiticity and positivity of $\dm(t)$ and thus respect the usual probability interpretation. 
In derivations of the Lindblad equation a couple of approximations must be made, such as assuming a weak system-bath interaction or the Born-Markov approximation (see, e.g., Refs.\ \onlinecite{Car93,BP10b,XTGP19}). 

Hence, we will merely use the master-equation approach to construct absorbing BC, i.e., the different approximations are controlled by choosing a setup where the central region of interest, which is initially excited by a local perturbation, is surrounded by a sufficiently large core region and finally by a boundary region where local Lindblad operators couple to the bath degrees of freedom and which must be large enough to fully absorb excitations emitted from the central part.
If perfectly absorbing BC can be constructed, one may in fact obtain the exact relaxation dynamics in the central part.

A similar idea has been applied recently \cite{AKvdL13} to compute steady-state properties of strongly correlated electron systems out of equilibrium. 
The required numerical solution of the Lindblad equation for interacting impurity systems can be carried out, e.g., with an exact-diagonalization approach in the superfermion representation of the Lindbladian \cite{DK11}. 
This requires auxiliary degrees of freedom and thus enlarges the Hilbert space, which, due to the two-body (Coulomb) interaction terms, is large anyway, such that the numerical implementation of Lindblad-type absorbing BC can become quite demanding in practice. 
For one-dimensional and impurity systems, density-matrix renormalization-group techniques are very powerful \cite{VGRC04,ZV04,PZ09}.

Actually, the Lindblad approach to absorbing BC appears to be perfectly suited for impurity models, where classical degrees of freedom are coupled to an {\em uncorrelated} electron system.
With the present study we focus on a system consisting of a single classical spin coupled to non-interacting conduction electrons with the goal to further develop the idea of absorbing BC.
We will demonstrate that the Lindblad approach can straightforwardly be adapted to the noninteracting case.
Surprisingly, however, we find that the resulting absorbing BC are not useful as demonstrated by comparing with results for open BC obtained for short propagation times.
While the coupling to the bath is found to almost perfectly suppress the unwanted reflections from the system boundaries, standard choices for the Lindblad parameters also {\em induce} unwanted artifacts, namely excitations {\em generated} initially at the boundaries, which are then propagating towards the core system and interfering with the physical dynamics. 
We therefore suggest to extend the Lindblad theory by considering Lindblad parameter {\em matrices} and by fixing those parameters such that a perfect suppression of the mentioned artificial initial excitations is achieved.
This requires to adapt the parameters to the system's initial state.
It is demonstrated that this approach leads to convincing results. 

The paper is organized as follows:
The following section \ref{sec:mod} introduces the model and the fundamental equations of motion. 
Sec.\ \ref{sec:absorb} discusses the standard Lindblad approach to absorbing BC and demonstrates its limitations. 
These are overcome with the novel BC introduced in Sec.\ \ref{sec:imp}. 
In Sec.\ \ref{sec:long} we discuss results demonstrating the progress made, and the conclusions are given in Sec.\ \ref{sec:con}.

\section{Model and equations of motion}
\label{sec:mod}

\begin{figure}[t]
\includegraphics[width=0.95\columnwidth]{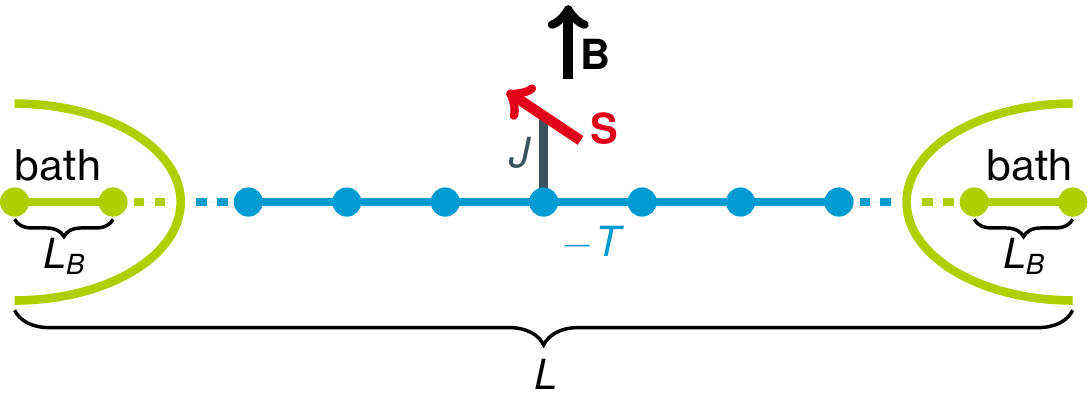}
\caption{
Sketch of the system geometry: 
A classical spin $\bm{S}$ of length $|\ff S|=\frac12$ is coupled via a local antiferromagnetic exchange interaction $J$ to a noninteracting system of electrons on a chain of $L$ sites. 
The hopping between nearest-neighboring sites is $-T$.
$L_{B}$ sites on the left and $L_{B}$ sites on the right edge are coupled to a bath.
The spin is located at the chain center and subjected to a local magnetic field $\ff B$.
Suddenly flipping the field direction induces the real-time dynamics.
}
\label{fig:sys}
\end{figure}

The generic model to discuss spin-relaxation dynamics is the $s$-$d$ exchange model \cite{VZ} where the spin $\ff S = (S_{x}, S_{y}, S_{z}) = \ff S(t)$ is treated as a classical dynamical variable, i.e., as a classical vector of fixed length $S = \frac12$. 
The spin is coupled to a system of noninteracting conduction electrons via a local antiferromagnetic exchange interaction.
The electron system serves as a large reservoir for the dissipation of energy and spin.
It is specified by the hopping $T_{ij}$ between the sites $i,j=1,...,L$ of a chain consisting of $L$ sites.
Throughout the study we consider hopping $T_{ij} = - T$ with $T>0$ between nearest neighbors $i$ and $j$ only. 
We assume half-filling with $N=L$ electrons in an isolated system with open boundary conditions (open BC).
Half-filling is also maintained when introducing a coupling of the sites close to the chain edges to an external bath in Sec.\ \ref{sec:absorb}.
Fig.\ \ref{fig:sys} provides a sketch of the system. 
The corresponding Hamiltonian (with open BC) reads
\be 
  H = \sum_{ij\sigma} T_{ij} c^\dagger_{i\sigma} c_{j\sigma} + J \bm{S} \bm{s}_{i_0} - \bm{S} \bm{B} \: .
\label{eq:ham}
\ee
Here, $c_{j\sigma}$ annihilates an electron at site $j$ with spin projection $\sigma = \uparrow,\downarrow$. 
The classical spin couples locally with strength $J>0$ to the local spin of the electron system, $\bm{s}_{i_0} = \frac12 \sum_{\sigma \sigma'} c^\dagger_{i_0\sigma} \bm{\tau}_{\sigma \sigma'} c_{i_0\sigma'}$, at site $i_{0}$ of the chain, where $\ff \tau = (\tau_{x}, \tau_{y}, \tau_{z})$ is a vector whose components are the Pauli spin matrices.
Furthermore, the model includes an external local magnetic field $\ff B$, which can be used to drive the classical spin.
Note that this does not couple to the electronic degrees for freedom.
The energy scale and (with $\hbar \equiv 1$) the time scale is set by choosing $T=1$.

Since the electron system is noninteracting, Wick's theorem applies, and all correlation functions factorize into one-particle correlations. 
A closed system of equations of motion, 
\be
  \frac{d}{dt} \bm{S}(t) = J \langle \bm{s}_{i_0} \rangle_t \times \bm{S}(t) - \bm{B} \times \bm{S}(t)
\label{eq:eoms}
\ee
and
\be
i \frac{d}{dt} \ff \rho(t) =  [\ff T_{\rm eff}(t), \ff \rho(t)]  \: , 
\label{eq:eomr}
\ee
can be obtained for the classical spin $\ff S=\ff S(t)$ and for the one-particle reduced density matrix $\ff \rho = \ff \rho(t)$ with elements 
\be
\rho_{i\sigma i' \sigma'}(t) = \langle \Psi(t) |c^{\dagger}_{i'\sigma'} c_{i\sigma}| \Psi(t) \rangle
\label{eq:dens}
\ee
where $|\Psi(t)\rangle$ is the many-body quantum state of the electron system, where 
$\langle \bm{s}_{i_0} \rangle_t =  \langle \Psi(t) | \bm{s}_{i_0} | \Psi(t) \rangle = \frac{1}{2} \sum_{\sigma\sigma'} \ff \tau_{\sigma\sigma'} \rho_{i_{0}\sigma'i_{0}\sigma}$, and where the effective hopping matrix $\ff T_{\rm eff}$ in Eq.\ (\ref{eq:eomr}) is given by the elements:
\be
  T^{\rm (eff)}_{i\sigma i'\sigma'}(t) = T_{ii'} \delta_{\sigma\sigma'} + \delta_{ii'} \frac J2 \ff S(t) \ff \tau_{\sigma\sigma'} 
  \: , 
\label{eq:teff}
\ee
see Refs.\ \onlinecite{Elz12,SP15} for a derivation and further details.

Suppose that initially the system is in its ground state for a given external field direction $\ff B_{0}$.
The formal purpose of the field is twofold: 
First, it breaks the SO(3) degeneracy of the ground state. 
Second, it will be employed to initiate the real-time dynamics at time $t=0$, namely by suddenly switching the field direction: $\ff B_{0} \to \ff B$. 
This sudden switch causes a local excitation of the system in the vicinity of site $i_{0}$.
In the course of time, the system is expected to relax such that the ground state will be restored locally. 
This requires that conserved quantities, i.e., energy and spin, must be transported away from $i_{0}$ and is in fact seen in the numerical solution of the equations of motion (\ref{eq:eoms}) and (\ref{eq:eomr}): 
Excitations are emitted from $i_{0}$ and propagate ballistically at a velocity $v = \ca O(T)$ set by the nearest-neighbor hopping. 
Assuming that the spin couples to the middle of the chain, i.e., 
\be
i_{0} = (L+1)/2
\ee
for odd $L$, this implies that after a time $\sim L / v$, the emitted excitations have reached the system boundaries, have been reflected and, after back-propagation, interfere with the local dynamics in the vicinity of site $i_{0}$. 

To avoid this unwanted finite-size effect in a practical calculation, a sufficiently large system is required. 
If one is interested in tracing the time evolution of the spin from the instant of the initial excitation to the fully relaxed final state, a system size $L \sim v \tau = \ca O(T \tau)$ is required.
Here, $\tau$ is the spin relaxation time. 
For a metallic state with $v \approx 2T$ \cite{SP15}, complete spin relaxation could be observed in computations for chains as long as $L=\ca O(10^{3})$ sites, but only at comparatively strong fields $B=\ca O(T)$. 
At weaker $B$ or for insulating states, however, the spin-relaxation time is expected to be possibly several order of magnitudes longer.
Since the computational effort for the numerical solution of the equations of motion scales as $L^{3}$ for large systems, such time scales $\gg 10^{3} / T$ cannot be reached in practice with the present theoretical setup.

\section{Construction of absorbing boundaries}
\label{sec:absorb}

A major goal of the this study is to construct system boundaries, which absorb the outgoing excitations emitted from the chain center.
The boundaries shall prevent any reflections to avoid the unwanted interference with the time evolution of local observables close to the central site $i_{0}$, such that their real-time dynamics in a sufficiently large environment of $i_{0}$ is practically indistinguishable from the dynamics of an infinite system ($L\to \infty$).
To this end we couple the outermost $L_{B}$ sites on the left and on the right edge of the chain to a suitable bath, while the remaining $L - 2 L_{B}$ sites are left untouched.
Typically we take $L_{B} \ll L$.
The model is displayed schematically in Fig.\ \ref{fig:sys}.

As a suitable framework for the construction of the absorbing boundaries, we consider the Lindblad master equation \cite{Lin76,Pea12}
\begin{align}
  \frac{d}{dt} \dm(t) 
  = 
  - i [H,\dm(t)] + \sum_\mu \left(
  2L_\mu \dm(t)L^\dagger_\mu - \{L^\dagger_\mu L_\mu, \dm(t) \}
  \right) 
\label{eq:lind}
\end{align}
for the many-body statistical operator $\dm(t)$. 
This appears as an attractive approach to construct absorbing boundaries as it preserves fundamental properties of the statistical operator, namely for all times $t$ we have $\tr \dm(t) = 1$, $\dm(t)^{\dagger} = \dm(t)$, and $\dm(t)\ge 0$. 
In Eq.\ (\ref{eq:lind}) the first term on the right-hand side is the von-Neumann term describing the system's unperturbed dynamics while the second one models the coupling to an external  bath via Lindblad operators $L_\mu$.
Here $\{\cdot,\cdot \}$ stands for the anticommutator.

Typically, the Lindblad operators are non-Hermitian and local. 
Here, we choose $L_{\mu} = L^{(r)}_{i\sigma}$ with $r=1,2$ and furthermore
\be
  L^{(1)}_{i\sigma} = \sum_{i'\sigma'} \alpha^{(1)}_{i\sigma i'\sigma'} \, c_{i'\sigma'} \; , \quad 
  L^{(2)}_{i\sigma} = \sum_{i'\sigma'} \alpha^{(2)\ast}_{i\sigma i'\sigma'} \, c^\dagger_{i'\sigma'}
  \; ,
\ee
i.e., we consider arbitrary linear combinations of annihilators or creators, respectively.
With this choice, one introduces a large number of unknown parameters to the theory, even if one takes into account that the sums over $i'$ are restricted to those sites coupling to the bath.
We will later see how these parameters are fixed in satisfactory way. 
In standard calculations one typically employs $r$-independent and diagonal matrices $\alpha^{(r)}_{i\sigma i'\sigma'} \propto \delta_{ii'} \delta_{\sigma\sigma'}$ to keep the number of parameters at a reasonable level.

\begin{widetext}
For the present case of a non-interacting electron system, the Lindblad equation (\ref{eq:lind}) for the statistical operator $\dm(t)$ can be strongly simplified and reformulated as a $2L \times 2L$ matrix equation for the one-particle reduced density matrix $\ff \rho(t)$, see Eq.\ (\ref{eq:dens}).
This is easily achieved by multiplying Eq.\ (\ref{eq:lind}) with $c^\dagger_{i'\sigma'} c_{i\sigma}$ from the right, by taking the trace, and using that $\tr(\dm(t) c^\dagger_{i'\sigma'} c_{i\sigma}) = \rho_{i \sigma i'\sigma'}(t)$.
We first get
\ba
  \frac{d}{dt} \rho_{i\sigma i'\sigma'}(t) 
  & = &
  -i \tr \left( [H,\dm(t)] \, c^\dagger_{i'\sigma'} c_{i\sigma} \right)
  \nonumber \\
  & + & 
  \sum_{j\tau j'\tau' j'' \tau''}  
  \alpha^{(1)}_{j\tau j'\tau'} \tr \left( 
  2 c_{j'\tau'} \dm(t) c^\dagger_{j''\tau''} 
  c^\dagger_{i'\sigma'} c_{i\sigma} 
  - 
  \{ c^\dagger_{j''\tau''} c_{j'\tau'},\dm(t) \}  
  c^\dagger_{i'\sigma'} c_{i\sigma} 
  \right)
    \alpha^{(1)\ast}_{j\tau j''\tau''} 
  \nonumber \\
  &+ &
  \sum_{j\tau j'\tau' j'' \tau''}  
  \alpha^{(2)\ast}_{j\tau j'\tau'} \tr \left( 
  2 c^{\dagger}_{j'\tau'} \dm(t) c_{j''\tau''} 
  c^\dagger_{i'\sigma'} c_{i\sigma} 
  - 
  \{ c_{j''\tau''} c^{\dagger}_{j'\tau'},\dm(t) \}  
  c^\dagger_{i'\sigma'} c_{i\sigma} 
  \right)
    \alpha^{(2)}_{j\tau j''\tau''} 
    \: .
\ea
Exploiting the cyclic invariance of the trace and using $\tr (\dm(t) O) = \langle O \rangle_{t}$ for an operator $O$, we find:
\ba
 \frac{d}{dt} \rho_{i\sigma i'\sigma'}(t) &=&-i \sum_{j\tau}
 \left(
 T^{\rm (eff)}_{i\sigma j\tau}(t) \rho_{j\tau i'\sigma'}(t) - \rho_{i\sigma j\tau} (t) T^{\rm (eff)}_{j\tau i'\sigma'}(t) 
 \right) 
 \nonumber \\
   & + & 
  \sum_{j\tau j'\tau' j'' \tau''}  
  \alpha^{(1)}_{j\tau j'\tau'} \left( 
  2 \langle
  c^\dagger_{j''\tau''} c^\dagger_{i'\sigma'} c_{i\sigma} c_{j'\tau'} 
  \rangle
  - 
  \langle 
  c^\dagger_{j''\tau''} c_{j'\tau'} 
  c^\dagger_{i'\sigma'} c_{i\sigma} 
  \rangle
  - 
  \langle 
  c^\dagger_{i'\sigma'} c_{i\sigma} 
  c^\dagger_{j''\tau''} c_{j'\tau'} 
  \rangle
  \right)
  \alpha^{(1)\ast}_{j\tau j''\tau''} 
  \nonumber \\ 
 &+&
  \sum_{j\tau j'\tau' j'' \tau''}  
  \alpha^{(2)\ast}_{j\tau j'\tau'} \left( 
  2 \langle
  c_{j''\tau''} c^\dagger_{i'\sigma'} c_{i\sigma} c^\dagger_{j'\tau'} 
  \rangle
  - 
  \langle 
  c_{j''\tau''} c^\dagger_{j'\tau'} 
  c^\dagger_{i'\sigma'} c_{i\sigma} 
  \rangle
  - 
  \langle 
  c^\dagger_{i'\sigma'} c_{i\sigma} 
  c_{j''\tau''} c^\dagger_{j'\tau'} 
  \rangle
  \right)
  \alpha^{(2)}_{j\tau j''\tau''}  
 \: .
\ea
The first term on the right-hand side reproduces the equation of motion (\ref{eq:eomr}), while the remaining ones can be simplified using the standard Fermi anticommutator rules. 
This results in the following equation of motion:
\ba
  \frac{d}{dt} \rho_{i\sigma i'\sigma'}(t)
  &= &
  -i \sum_{j\tau} \left(
  T^{\rm (eff)}_{i\sigma j\tau}(t) \rho_{j\tau i'\sigma'}(t) - \rho_{i\sigma j\tau}(t) T^{\rm (eff)}_{j\tau i'\sigma'}(t)
  \right)
  \nonumber \\
  & - &
  \sum_{j\tau j' \tau'}  
  \alpha^{(1)}_{j\tau i'\sigma'} 
  \rho_{i\sigma j'\tau'} 
  \alpha^{(1)\ast}_{j\tau j'\tau'} 
  - 
  \sum_{j\tau j'\tau'}  
  \alpha^{(1)}_{j\tau j'\tau'} 
  \rho_{j'\tau'i'\sigma'} 
  \alpha^{(1)\ast}_{j\tau i\sigma} 
  \nonumber \\
   &-&
  \sum_{j\tau j' \tau'}  
  \alpha^{(2)\ast}_{j\tau i\sigma} 
  \rho_{j'\tau' i'\sigma'} 
  \alpha^{(2)}_{j\tau j'\tau'}  
  - 
  \sum_{j\tau j'\tau'}  
  \alpha^{(2)\ast}_{j\tau j'\tau'} 
  \rho_{i\sigma j'\tau'} 
  \alpha^{(2)}_{j\tau i'\sigma'}  
  +
  2\sum_{j\tau}  
  \alpha^{(2)\ast}_{j\tau i\sigma} 
  \alpha^{(2)}_{j\tau i'\sigma'}
  \; , 
\ea
which can be written in matrix form:
\ba
  \frac{d}{dt} \ff \rho(t)
  &= &
  -i [ \ff T_{\rm eff}(t) , \ff \rho(t) ]
  -
  \ff \rho(t) \ff \alpha_{1}^{\dagger} \ff \alpha_{1}
  - 
  \ff \alpha_{1}^{\dagger} \ff \alpha_{1} \ff \rho(t) 
  -
  \ff \alpha_{2}^{\dagger} \ff \alpha_{2} \ff \rho(t) 
  - 
  \ff \rho(t) \ff \alpha_{2}^{\dagger} \ff \alpha_{2}
  + 2 \ff \alpha_{2}^{\dagger} \ff \alpha_{2}
  \: .
\label{eq:eomrc}
\ea
We define the Hermitian and nonnegative matrices
\be
  \ff \gamma = \ff \alpha_{1}^{\dagger} \ff \alpha_{1}
  +
  \ff \alpha_{2}^{\dagger} \ff \alpha_{2}
  \; , \quad
  \ff \Gamma = \ff \alpha_{2}^{\dagger} \ff \alpha_{2}
  \: ,
\label{eq:nonneg}
\ee
such that the equation reads as
\ba
  \frac{d}{dt} \ff \rho(t) 
  =
  -i [ \ff T_{\rm eff}(t) , \ff \rho(t) ]
  -
  \{\ff \gamma, \ff \rho(t)\}
  + 2\ff \Gamma
  \: .
\label{eq:eomd}
\ea
This replaces Eq.\ (\ref{eq:eomr}). 
Note that the effective hopping matrix depends on $\ff S(t)$, and thus Eq.\ (\ref{eq:eomd}) must still be supplemented by the equation of motion (\ref{eq:eoms}) for the classical spin.
\end{widetext}

Eqs.\ (\ref{eq:eoms}) and (\ref{eq:eomd}) describe the relaxation of the system after an initial excitation of the localized spin.
In the core system, i.e., for $L_{\rm B} < i < L+1-L_{\rm B}$, conservation laws hold locally. 
Hence, energy, spin and particles are transported to the chain edges and dissipated to the external baths for  finite Lindblad coupling parameters $\ff \Gamma, \ff \gamma$.
The Lindblad parameters are taken to be nonzero at the boundaries only.

To test the quality of the absorbing boundaries implemented with the standard Lindblad equation and generic Lindblad paramters, we consider a manifestly particle-hole symmetric electron system at half-filling, i.e., $\sum_{\sigma} \rho_{i\sigma i \sigma}(t)=1$. 
For the sake of simplicity, we assume diagonal coefficient matrices $\ff \alpha_{r}$ with real spin- and $r$-independent diagonal elements: 
\be
\alpha^{(r)}_{i\sigma i'\sigma'} = \delta_{ii'}\delta_{\sigma\sigma'} \alpha_{i} \: . 
\label{eq:stan}
\ee
This implies $\ff \gamma = 2 \ff \Gamma$ and $\Gamma_{i\sigma i'\sigma'} = \delta_{ii'} \delta_{\sigma\sigma'} \Gamma_{i}$.
With this standard choice, particle-number conservation is maintained as is easily verified by taking the trace of both sides of Eq.\ (\ref{eq:eomd}) and noting that $\langle N \rangle = \tr \ff \rho(t)$.
We furthermore set the parameters either as constant, 
\be
 \Gamma_i = \Gamma > 0\: , 
\label{eq:const}
\ee
for all sites coupling to the external bath, or choose them to increase linearly with increasing distance to the outermost sites of the core system,
\be
 \Gamma_{i} =
 \begin{cases}
  (L_B +1 -i) \Gamma_\text{min} & i \leq L_B \\
  0 & L_B < i < L+1-L_B \\
  (i-(L-L_B))\Gamma_\text{min} & i \geq L+1-L_B
 \end{cases}
 \: ,
\label{eq:lin} 
\ee
with $\Gamma_{\rm min}>0$, and use $\Gamma$ or $\Gamma_{\rm min}$ to optimize the absorbing properties of the coupling to the bath.

\begin{figure}[t]
\includegraphics[width=0.9\columnwidth]{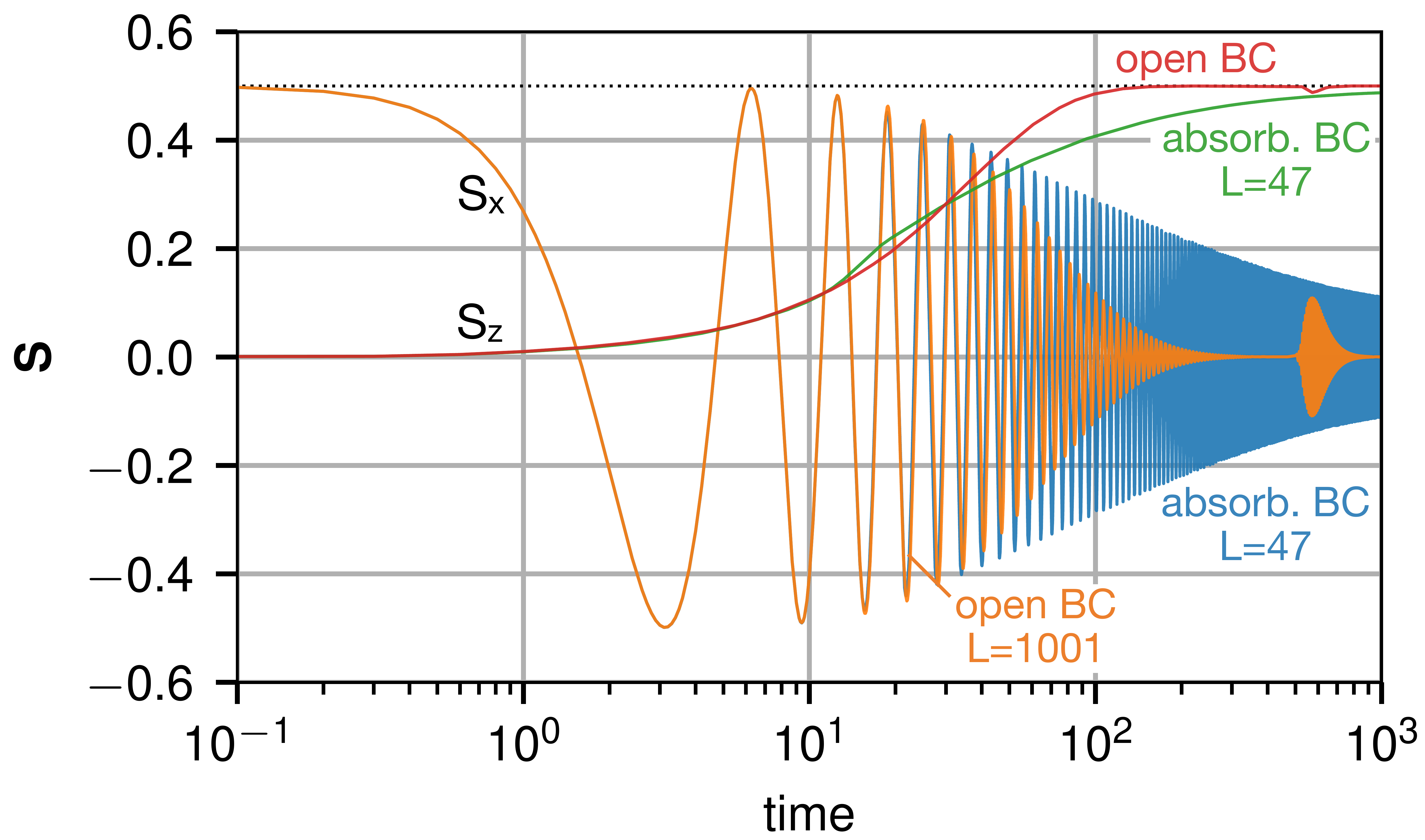}
\caption{
Time evolution of the $z$- and the $x$-component of the classical spin coupled to an electron system with  n.n.\ hopping $-T$ at half-filling after a sudden switch of the local magnetic field from $x$ to $z$ direction (see text for details).
Red/orange lines: Standard theory for a chain with open boundary conditions (open BC) with $L=1001$ sites ($i_{0}=501$, $J=1$, $B=1$).
Green/blue lines:  Calculation with absorbing boundaries (absorbing BC) [Eqs.\ (\ref{eq:eoms}), (\ref{eq:eomd}), (\ref{eq:stan}) and (\ref{eq:lin})] for $L=47$ ($i_{0}=24$, $J=1$, $B=1$, $L_{\rm B}=5$, $\Gamma_\text{min}=0.2$).
Energy and time scales set by $T=1$, $\hbar=1$.
}
\label{fig:comp1}
\end{figure}

To check the effect of absorbing boundaries, we compare numerical results obtained with the standard theory for a large system ($L=1001$) and open BC to results obtained with Eq.\ (\ref{eq:eomd}) for a much smaller system ($L=47$) and absorbing BC, see Fig.\ \ref{fig:comp1}. 
For the integration of the equations of motion a high-order Runge-Kutta technique with variable step size is employed. 
We set $J=1$ and $B=1$, as we expect a comparatively short spin-relaxation time $\tau$ for this choice of model parameters. 
The local magnetic field is suddenly switched from $x$- to $z$-direction to initiate the dynamics, i.e., we prepare the system in its ground state for $\ff B_{0}$ pointing in $x$-direction by diagonalization of the effective hopping matrix and by filling the effective one-particle eigenstates up to the Fermi level to reach half-filling. 
For the subsequent dynamics starting at $t=0$, the field $\ff B$ points into the $z$-direction.

In the case of open BC, the $x$-component of the classical spin immediately starts to oscillate (see Fig.\ \ref{fig:comp1}).
Together with the $y$-component (not displayed) this just reflects the Larmor precession of the spin around the field direction. 
The precession frequency is $\omega \approx B$. 
Looking at the $z$-component we see that the spin relaxes to the new field direction on a time scale of $t \approx 200$.
Our physical expectation is that after reaching its new ground-state direction, the spin dynamics should basically stop. 
As can be seen in Fig.\ \ref{fig:comp1}, however, there is an unphysical revival of the dynamics for $t \gtrsim 500$. 
Further revivals at still later times are expected as well.
These are in fact caused by the effect of excitations reaching the site $i_{0}$ after back reflection from the system boundaries. 
The time scale for this unwanted artifact is approximately given by twice the distance of $i_{0}$ to the edges of the system size, $2 \cdot L/2 \approx 1000$, divided by the the Fermi velocity $v_{\rm F}=2$. 

Let us now compare with the results obtained for the small system ($L=47$) with absorbing BC. 
We employ the model with linearly increasing coupling parameters, Eq.\ (\ref{eq:lin}), starting with $\Gamma_{\rm min} = 0.2$ and use $L_{\rm B}=5$ absorbing sites on each edge, such that the core system has $L-2L_{\rm B} = 37$ sites. 
We find that, initially, up to about $t=10$, the dynamics is reproduced more or less correctly.
For $t<10$, there are tiny deviations, which are most clearly seen in the $z$-component of the spin. 
These could be attributed, e.g., to the coarser description of the initial Fermi-sea ground state. 
The main effect for $t\gtrsim 10$, however, appears to be again related to the presence of the boundaries as becomes obvious when comparing calculations for different system sizes $L$ (not displayed).
Compared to the results for open BC, these deviations must obviously show up much earlier, at about $t=23$, due to the much shorter distance to the edges ($L=47$ vs.\ $L=1001$).
We find, however, that they come even earlier by about a factor of two.

At later times $t \gtrsim 100$, the predicted dynamics deviates strongly and full spin relaxation, if present at all, is massively delayed with $\tau \gg 1000$. 
We conclude that absorbing BC, naively derived from the Lindblad approach with a standard parameter choice, lead to an unacceptable impact on the spin (and electron) dynamics. 
Note, however, that there are in fact no visible effects, which hint to {\em reflections} from the boundaries.
Hence, the presently discussed absorbing BC do absorb the outgoing excitations, but at the same time strongly disturb the time evolution.
Let us point out that this does not depend very much on the parameter choice as has been checked by varying $\Gamma_{\rm min}$ and $L_{B}$. 
Also for spatially constant parameters, see Eq.\ (\ref{eq:const}), the results do not improve or get worse  significantly.

\begin{figure}[t]
\includegraphics[width=0.8\columnwidth]{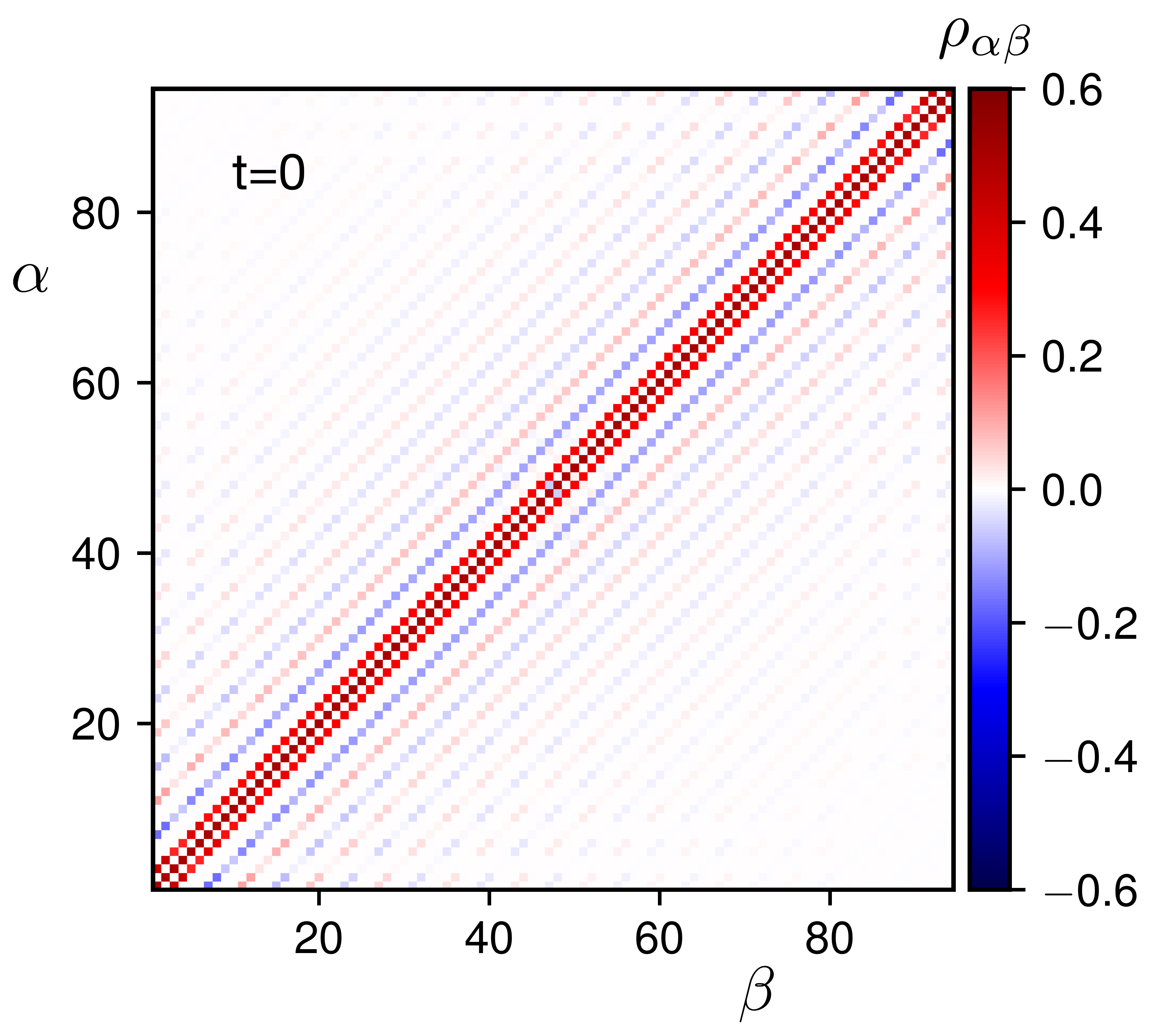}
\caption{
Initial one-particle reduced density matrix at time $t=0$ for a system with $L=47$ sites, open BC, and the impurity spin at the central site $i_{0} = (L+1)/2$ pointing in $x$-direction. 
The color coding is indicated by the bar on the right side.
Exchange coupling $J=1$.
We display the elements $\rho_{\alpha\beta}$ of $\ff \rho$ using the combined site-spin (``orbital'') index $\alpha \equiv 2 i - \frac{1}{2} (1+z_\sigma) = 1, ..., 2L$ with $z_{\uparrow} = +1$, $z_{\downarrow}= -1$.
}
\label{fig:rho0}
\end{figure}

Our strategy in the following is to find the cause of the problem and to modify the absorbing boundary conditions accordingly. 
Fig.\ \ref{fig:rho0} displays the initial one-particle reduced density matrix at time $t=0$.
The density matrix at time $t=0$ is constructed as the ground-state density matrix for $\ff B_{0}= \ff e_{x}$, i.e., for the classical spin pointing in $x$-direction.
Since $J>0$, the electron magnetic moment at $i_{0}$ is antiferromagnetically oriented.
We see that $\rho_{i\sigma i\sigma}=0.5$ for all sites, corresponding to half-filling. 
Further, $\rho_{i\uparrow i\downarrow}=\rho_{i\downarrow i\uparrow}$ for an $x$-polarized state.
The site off-diagonal elements $\rho_{i\sigma i'\sigma}$ with $i \ne i'$show a damped oscillation with increasing distance $|i-i'|$. 
Close to $i_{0}$ and particularly close to the chain edges, there are some Friedel-like oscillations of the diagonal elements $\rho_{i\sigma i\sigma}$ as function of $i$.
The oscillations induced by the edges are strongly damped, such that the density-matrix elements close to the center are essentially unaffected.

\begin{figure*}[t]
\includegraphics[width=1.95\columnwidth]{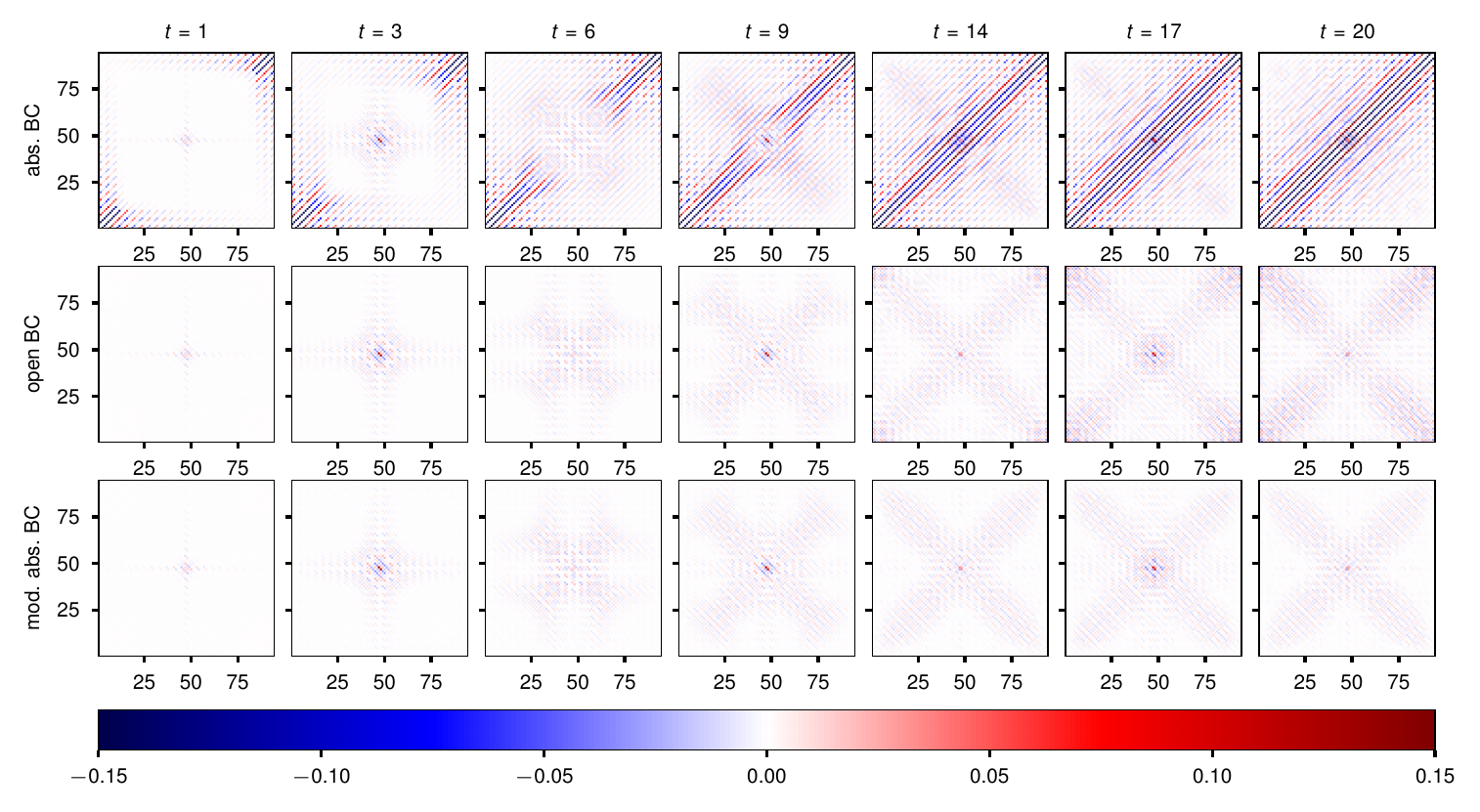}
\caption{
Time dependence of the one-particle reduced density matrix for a system of $L=47$ sites.
The color code (see bottom) quantifies the real part of the difference to the initial density matrix, $\mbox{Re} (\ff \rho(t) - \ff \rho(0))$, at selected instants of time, see the time labels at the top.
Representation of the elements $\rho_{\alpha\beta}$ as in Fig.\ \ref{fig:rho0} using the orbital index $\alpha = 2 i - \frac{1}{2} (1+z_\sigma) = 1, ..., 2L$.
{\em Middle panel:} system with open BC.
{\em Upper panel:} same system but with absorbing BC based on the standard Lindblad approach
[Eqs.\ (\ref{eq:eomd}), (\ref{eq:stan}) and (\ref{eq:lin})]. 
{\em Lower panel:} same system but with modified novel absorbing BC (see text).
Other parameters as in Fig.\ \ref{fig:comp1} or Fig.\ \ref{fig:spin} respectively. 
}
\label{fig:rhot}
\end{figure*}

Fig.\ \ref{fig:rhot} shows the time evolution of the density-matrix elements for a system with $L=47$ sites.
As compared to the initial density matrix $\ff \rho(0)$ the time-dependent {\em deviations} of the matrix elements, $\ff \rho(t) - \ff \rho(0)$, are typically smaller by more than an order of magnitude (note the different scales encoded with the color plots in Figs.\ \ref{fig:rho0} and \ref{fig:rhot}).
Hence, only (the real part of) the difference is plotted. 
For open BC (middle panel of Fig.\ \ref{fig:rhot}) we see an overall oscillation of elements $\rho_{i\sigma i'\sigma'}$ with $i,i'$ close to $i_{0}$ (central site) with a period approximately given by $2\pi / \omega_{\rm L}$, where $\omega_{\rm L} = B = 1$ is the Larmor frequency. 
More important, however, one finds spin-dependent excitations being emitted from the central region. 
These oscillate with the same frequency but are phase shifted depending on the distance to $i_{0}$, i.e., we see a propagation of a wave packet through the lattice. 
This propagation is found to be equally pronounced for the spatially diagonal ($i=i'$) elements of $\rho_{i\sigma i' \sigma'}$ as well as for the off-diagonal ones.
At later times $t$, approximately given by the distance $L/2$ divided by the the Fermi velocity $v_{\rm F}=2$, i.e., $t \gtrsim 10$, the excitations reach the edges, are back-reflected and, for still later times, lead to the unwanted interference with the relaxation dynamics close to $i_{0}$.

For the same system but with absorbing BC based on the Lindblad approach with standard parameter choice, Eqs.\ (\ref{eq:eomd}), (\ref{eq:stan}) and (\ref{eq:lin}), there are several defects that are uncovered with the upper panel of Fig.\ \ref{fig:rhot}. 
First, the comparison of results for open and absorbing BC at early times shows that the presence of the coupling to the bath {\em induces} artificial excitations, which {\em start} close to the edges and propagate to the central region with Fermi velocity and finally, at times $\approx (L/2)/v_{\rm F}$, interfere with the spin-relaxation dynamics close to $i_{0}$.
This actually explains the different time evolution of the classical spin in Fig.\ \ref{fig:comp1} for times $t \gtrsim (L/2)/v_{\rm F} \approx 12$.
This artifact stems from bath contributions to the equations of motion, which are nonzero in the initial state at $t=0$ and must be avoided by an improved model for the coupling to the bath.

Second, as a consequence of the damping terms in the equation of motion (\ref{eq:eomd}) for the one-particle reduced density matrix, we see that all its nondiagonal elements $i \ne i'$ are exponentially approaching zero.
In the full dynamics, on the other hand, this is not the case at all. 
Especially the elements with $i' = i \pm 1$, have a considerable absolute magnitude at $t=0$ (Fig.\ \ref{fig:rho0}), and essentially do not decrease in the course of time. 

Finally, absorbing BC based on the standard Lindblad approach do not introduce absorption of excitations propagating along the antidiagonal of the density matrix. 
Such excitations on the antidiagonal, however, are clearly seen in the middle panel of Fig.\ \ref{fig:rhot} and are actually of the same order of magnitude as compared to the diagonal.
Hence, absorption of both, diagonal and antidiagonal excitations reaching the edges, must be included in a modified coupling to the bath.

\section{Improved absorbing boundaries}
\label{sec:imp}

To analyze their origin and to remove the artifacts, we first consider the equation of motion (\ref{eq:eomd}) at time $t=0$. 
For a quench of the magnetic-field direction, the density matrix $\ff \rho(t)$ commutes with the effective hopping matrix $\ff T_{\rm eff}(t)$ at $t=0$. 
For an infinite system or for a system with open boundaries, this would imply $d \ff \rho(t) / dt |_{t=0} = 0$. 
Note that there is a finite torque on the local impurity spin that initiates the dynamics, and the updated impurity-spin direction will impact $\ff \rho(t)$ for $t>0$. 
With standard Lindblad boundaries, however, there is a nonzero time derivative of $\ff \rho(t)$ already at $t=0$:
\be
  \frac{d}{dt} \ff \rho(t) |_{t=0} =  - \{ \ff \gamma , \ff \rho(0) \} + 2 \ff \Gamma 
  \: , 
\label{eq:initialbath}
\ee
which gives rise to dynamics due to the mere presence of the bath and which starts from the system boundaries.
Avoiding this artificial cause of dynamics implies the following condition on the Lindblad parameters:
\be 
\ff \Gamma  = \frac{1}{2} \{ \ff \gamma , \ff \rho(0) \} \; , 
\label{eq:cond}
\ee
i.e., we must necessarily choose the parameters dependent on the initial system state.
Furthermore, this condition also implies an $r$-dependent choice of the coefficient matrices $\ff \alpha_{r}$, see Eq.\ (\ref{eq:nonneg}).
Using Eq.\ (\ref{eq:cond}) to eliminate $\ff \Gamma$, the resulting equation of motion reads:
\ba
  \frac{d}{dt} \ff \rho(t) 
  =
  -i [ \ff T_{\rm eff}(t) , \ff \rho(t) ]
  -
  \{\ff \gamma, \ff \rho(t) - \ff \rho(0) \}
  \: .
\label{eq:eomm}
\ea
We emphasize that all properties that are constitutive for the general Lindblad approach apply to this equation as well, as it exactly derives from the fundamental Lindblad equation (\ref{eq:lind}) by merely specializing to a noninteracting electron system and by a special parameter choice only. 
Particularly, Eq.\ (\ref{eq:eomm}) therefore respects the Hermiticity and the nonnegativity of $\ff \rho(t)$ at all times $t$. 

However, there are restrictions for the choice of the parameter  $\ff \gamma$, which must be taken care of. 
To discuss this, let us first construct the general formal solution of Eq.\ (\ref{eq:eomm}), assuming that the impurity spin $\ff S(t)$ and thus the time-dependence of $\ff T_{\rm eff}(t)$ is given.
Eq.\ (\ref{eq:eomm}) represents a linear inhomogenous system of first-order ordinary differential equations. 
The corresponding homogeneous system, 
$\frac{d}{dt} \ff \rho(t) 
  =
  -i [ \ff T_{\rm eff}(t) , \ff \rho(t) ]
  -
  \{\ff \gamma, \ff \rho(t) \}
$,
can be written as $i (d/dt) \ff \rho = \ff \Theta \ff \rho - \ff \rho \ff \Theta^{\dagger}$ with $\ff \Theta \equiv \ff T - i \ff \gamma$ and is thus solved by $\ff \rho = \ff U \ff \rho_{0} \ff U^{\dagger}$ for the initial condition $\ff \rho(t=0) = \ff \rho_{0}$. 
Here, $\ff U =\ff U(t) = \ff U(t,0)$ with $\ff U(t,t') = \ca T \exp (- i \int_{t'}^{t} d\tau \ff \Theta(\tau))$ (for $t>t'$) is a {\em nonunitary} time-evolution matrix formally constructed with the help of the time-ordering operation $\ca T$.
A special solution of the inhomogeneous system is easily obtained with the ansatz $\ff \rho = \ff U \widetilde {\ff \rho} \ff U^{\dagger}$.
We find $\dot{\widetilde {\ff \rho}} = \ff U^{-1} \{\ff \gamma , \ff \rho_{0} \} \ff U^{\dagger -1}$. 
The desired special solution with initial condition $\widetilde {\ff \rho}(t=0) = 0$ is obtained by integration and back transformation from $\widetilde {\ff \rho}$ to $\ff \rho$. 
Adding the solution of the homogeneous system, we finally obtain:
\ba
  \ff \rho(t) 
  &=& 
  \ff U(t,0) \ff \rho(0) \ff U(t,0)^{\dagger}
\nonumber \\  
   &+&
  \int_{0}^{t} d \tau \,
  \ff U(t,\tau)
  \{\ff \gamma , \ff \rho(0) \} 
  \: 
  \ff U(t,\tau)^{\dagger}
  \: .
\ea
Note that for finite damping $\ff \gamma$ the backwards time evolution $\ff U(t,t')^{-1} = \ff U(t',t) = \widetilde{\ca T} \exp (- i \int_{t}^{t'} d\tau \ff \Theta(\tau))$ (for $t>t'$ and with the antichronological ordering $\widetilde{\ca T}$) is generally different from the adjoint of the time evolution $\ff U(t,t')^{\dagger} \ne \ff U(t',t)$.
Due to the nonunitarity of $\ff U$, damping is not only described by the second term including a memory effect but also by the first one.

One immediately sees that $\ff \rho(t)$ is Hermitian and nonnegative for all $t$, if (i) the anticommutator $\{\ff \gamma , \ff \rho(0) \}$ is nonnegative, and if (ii) $\ff \gamma$ is Hermitian.
Furthermore, we must have (iii) $\ff \gamma \ge 0$ to ensure that the first ``homogeneous'' term remains bounded for $t \to \infty$. 
The conditions (i) and (iii) are also obvious from Eqs.\ (\ref{eq:nonneg}) and (\ref{eq:cond}).

All conditions (i) - (iii) can be satisfied as follows:
We diagonalize the initial density matrix, $\ff \rho(0) = \ff V^{\dagger} \ff n \ff V$, with a unitary matrix $\ff V$.  
The elements of the diagonal matrix $\ff n$, the natural occupations, are nonnegative since $\ff \rho(0) \ge 0$. 
The rows of $\ff V$ are the corresponding natural orbitals.
Note that, for an infinite and translationally invariant system, the natural orbitals are delocalized states and labelled by a wave vector. 
Hence, for a finite but large $L$ we expect them to be rather delocalized as well. 
Using $\ff V$, we can now define $\ff \gamma \equiv \ff V^{\dagger} \ff g \ff V$, where $\ff g$ is a real, nonnegative and diagonal matrix. 
With this choice, we immediately have $\ff \gamma^{\dagger} = \ff \gamma$ and $\ff \gamma \ge 0$, i.e., conditions (ii) and (iii) are satisfied. 
Furthermore, since $\ff \gamma$ and $\ff \rho(0)$ are, by construction, simultaneously diagonalized by the same unitary transformation $\ff V$, they must commute.
This immediately implies condition (i). 
The remaining degrees of freedom, the elements of the diagonal matrix $\ff g$, should be used to localize $\ff \gamma$ close to the system boundary. 
Strictly speaking, we need to satisfy $\ca O((L-L_{B})^{2})$ conditions of the form $\gamma_{ii'\sigma\sigma'} = 0$ for $i,i'$ in the core system, having only $\ca O(L)$ parameters at our disposal.
While this is not an obstacle in principle, it would imply that the boundary region with finite coupling to the bath extends over almost the whole system and that the remaining core system is comparatively small.
From a computational point of view this is highly inconvenient.

In practice, it has turned out, however, that a more pragmatic and much simpler procedure is fully satisfying. 
We take $\ff \gamma$ as diagonal right from the start and set $\gamma_{i\sigma} = \gamma$ with $\gamma >0$ for a small number of sites $2L_{B}$ coupling to the external bath and $\gamma_{i\sigma}=0$ else. 
Alternatively, a linear $\gamma$-profile, analogous to Eq.\ (\ref{eq:lin}) may be employed.
This implies that generically $\ff \gamma$ does not commute with $\ff \rho(0)$, and hence 
$2 \ff \Gamma = \{ \ff\gamma, \ff \rho(0) \}$, see Eq.\ (\ref{eq:cond}), may develop negative eigenvalues. 
While there are negative eigenvalues of $2\ff \Gamma$ indeed, as is easily seen numerically, these have a small modulus for all cases studied and particularly for setups with a small boundary and a large core region, i.e., for the conceptually and computationally attractive case.
Causality problems, such as negative densities $\rho_{i\sigma i\sigma}<0$ have not been observed.
One may also relax the condition (\ref{eq:cond}) and replace the initial density matrix by the $J=0$ density matrix for the computation of $\ff \Gamma$, with the idea to work with a spin-independent $\ff \Gamma$ matrix.
Again, this is unproblematic in practice, as the finite coupling to the classical spin does not affect the density-matrix elements in the boundary region substantially if $L$ is reasonably large.

\begin{figure}[t]
\includegraphics[width=0.9\columnwidth]{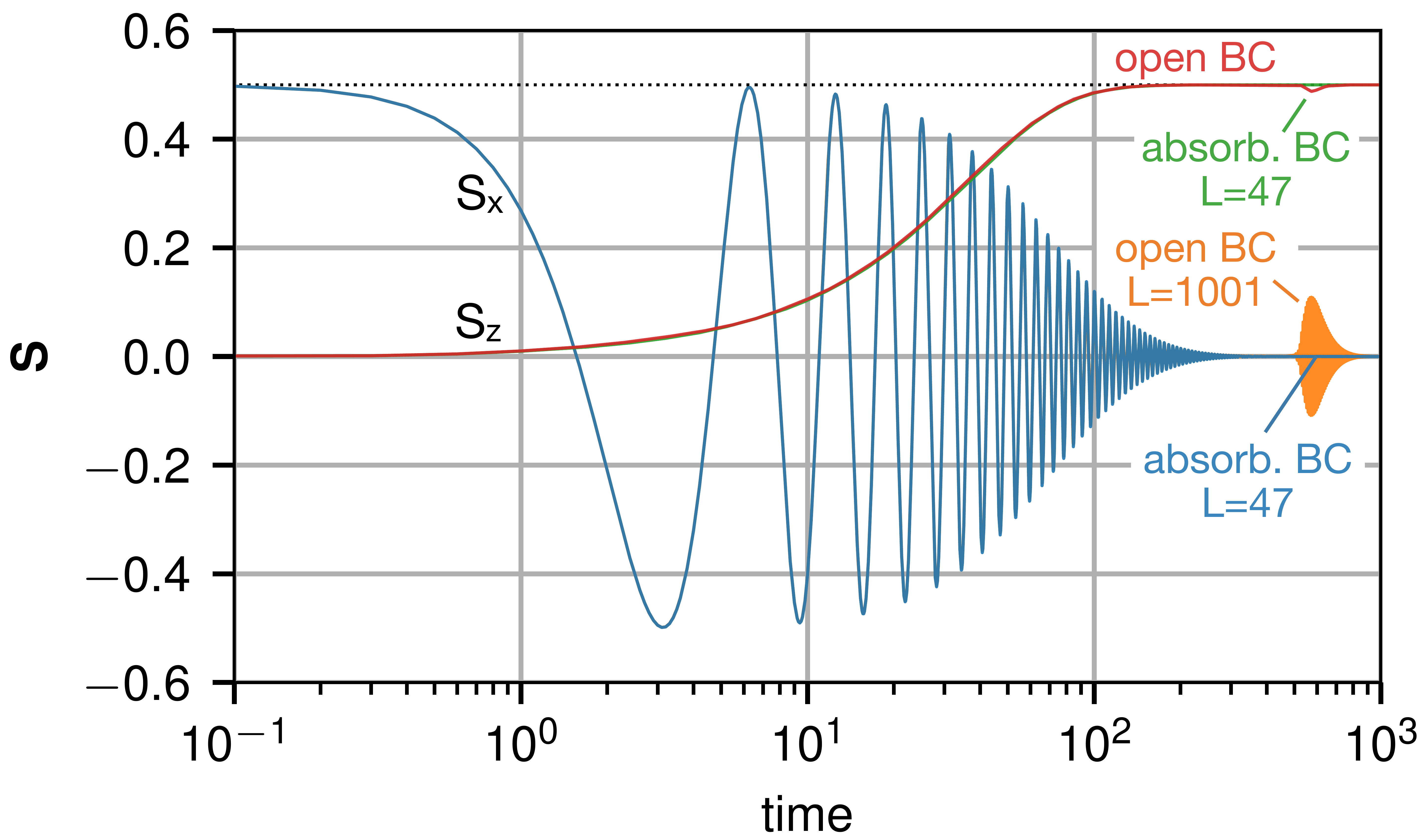}
\caption{
Time evolution of the $z$- and the $x$-component of the classical spin as in Fig.\ \ref{fig:comp1} but here the results of the standard theory (red/orange lines) for open BC and $L=1001$ are compared to those obtained for $L=47$ sites (green/blue lines) with modified novel BC (linear profile and $\gamma_\text{min}=0.2$). 
Other parameters as in Fig.\ \ref{fig:comp1}.
}

\label{fig:spin}
\end{figure}

To test the novel construction of absorbing BC, we solve the coupled system of Eqs.\ (\ref{eq:eoms}) and (\ref{eq:eomm}) for the comparatively small system with $L=47$ sites. 
The lower panel of Fig.\ \ref{fig:rhot} displays the time evolution of the one-particle reduced density matrix as obtained with the modified absorbing BC. 
Comparing with the results obtained for open BC (middle panel) at early instants of time ($t\le 9$) and in the central region for $i,i'$ close to $i_{0}$, only marginal differences are found, which are by far too small to be visible in the figure. 
In particular, all fine details of the spatial structure of the density matrix are reproduced correctly.

For later times, see $t=20$, for example, there are still no deviations in the central region.
This is as desired.
In the calculation with open boundaries, we expect unphysical interference effects only for times $t \gtrsim 2 i_{0} /v_{\rm F} \approx 23$.
Off the central region, however, artifacts start for $t=20$ and also for earlier times, e.g., $t=14$, but only for sites $i$ and $i'$ far from the central site $i_{0}$, both on the diagonal and the antidiagonal (see, e.g., the middle panel for $t=14$, around $i=1$, $i'=1$ and around $i=1$, $i'=L$).
On the other hand, the calculations with modified absorbing BC are entirely free from those artifacts.  
Comparing with the simple absorbing BC based on the naive application of the Lindblad approach (upper panel), demonstrates the progress made, in particular if one takes into account the fact the small scale of {\em differences} to the initial-state ($t=0$) density matrix. 

We conclude that the absorption of the outgoing excitations is perfectly accomplished with the novel approach, Eq.\ (\ref{eq:eomm}), and that therefore the temporal development of the density matrix in the physical core of the system indeed reflects the temporal development of the infinite system very accurately.

This is also nicely seen in the resulting relaxation dynamics of the classical spin. 
In Fig.\ \ref{fig:spin} we compare $\ff S(t)$ as obtained from the calculation for the small system with $L=47$ sites and with the new absorbing BC to corresponding results of a calculation with open BC but for a much larger system ($L=1001$).
For the chosen system parameters the spin relaxation time amounts to $\tau \approx 200$ inverse hoppings. 
We note that for $t \gtrsim \tau$ artifical interference with excitations back-reflected from the edges manifests itself in an unphysical revival of the dynamics starting at $t \approx 500$ inverse hoppings in the calculation done for open BC, while there is no such effect visible for modified absorbing BC.
For times shorter than $t\approx 500$, the agreement between the results obtained for $L=1001$ (open BC) and for $L=47$ sites (absorbing BC) is not perfect but extremely good, such that deviations are more or less invisible on the scale of the figure.
Remaining discrepancies can be eliminated systematically by increasing the core system size. 

\section{Accessing long time scales}
\label{sec:long}

\begin{figure}[t]
\centering
\includegraphics[width=0.8\columnwidth]{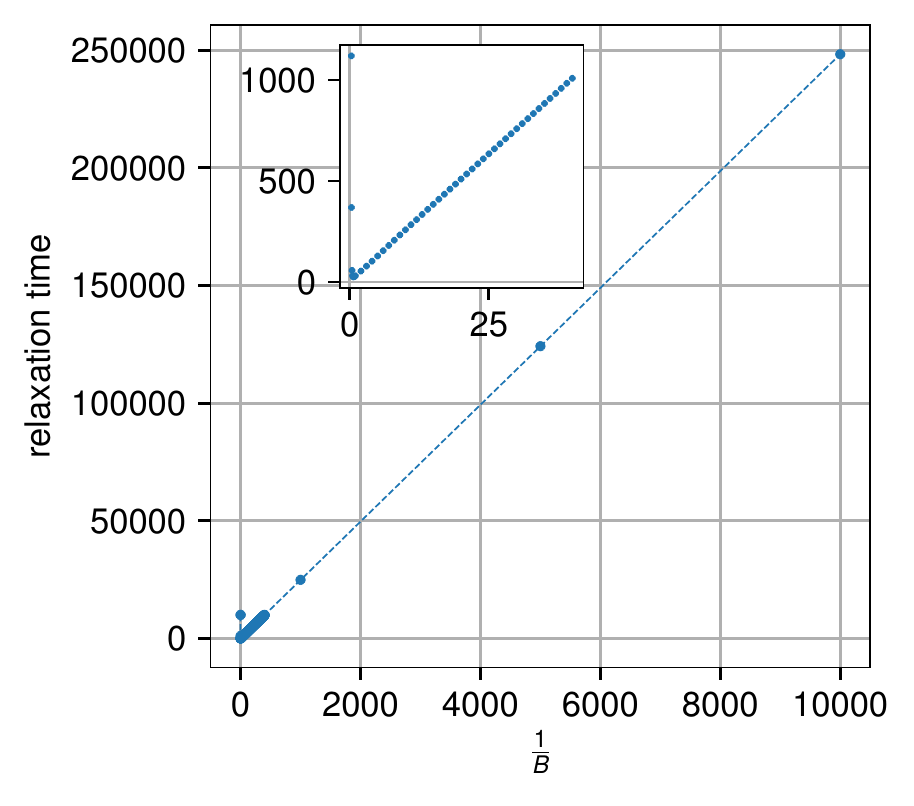}
\caption{
Relaxation time $\tau$ as a function of $1/B$. 
Calculations for $i_{0}=1$ (spin couples to the ``left'' edge), $L=46$, $J=1$, and modified novel BC for the ``right'' edge (linear profile, $\gamma_{\rm min} = 0.2$, $L_B=5$).
} 
\label{fig:met}
\end{figure}

The benefit of the novel absorbing BC is that much longer time scales are accessible. 
This is demonstrated with Fig.\ \ref{fig:met}, which displays the relaxation time $\tau$ as a function of the magnetic field strength $B$. 
For convenience the classical spin is coupled to the first site of the chain, $i_{0}=1$, and the absorbing BC are implemented, with $L_{B}=5$ sites coupling to the bath, for the opposite edge.
We define $\tau$ pragmatically as the time required for $S_{z}(t)$ to reach $95\%$ of its fully relaxed value $S_{z}(t\to \infty) = 0.5$.
As can be seen in the figure, for very weak fields, down to $B=1\cdot 10^{-4}$, the relaxation time approaches $\tau \approx 250,000$ in units of the inverse hopping parameter, i.e., the coupled microscopic real-time dynamics of the spin and the conduction-electron system can be traced on a time scale, which is by more than five orders of magnitude longer than the intrinsic bare time scale of the electron system that is set by the inverse hopping $1/T=1$.
This is way beyond what can be reached with conventional calculations using open BC.

It is instructive to compare the results with the prediction of the Landau-Lifschitz-Gilbert (LLG) approach \cite{llg},
\be
\tau \propto \frac{1+\alpha^{2}}{\alpha} \frac{1}{B} \: , 
\label{eq:taullg}
\ee
where $\alpha$ is the Gilbert damping parameter, see Ref.\ \cite{Kik56}.
Starting from the simplified model considered here, the LLG equation can be derived by lowest-order perturbation theory in $J$ and by a Markov approximation assuming that the spin dynamics is much slower than the electron dynamics, i.e., by assuming that the strength of the local field $B$ is weak on the scale given by the nearest-neighbor hopping (see, e.g., Ref.\ \cite{SP15} for a detailed discussion).

\begin{figure}[t]
\centering
\includegraphics[width=0.8\columnwidth]{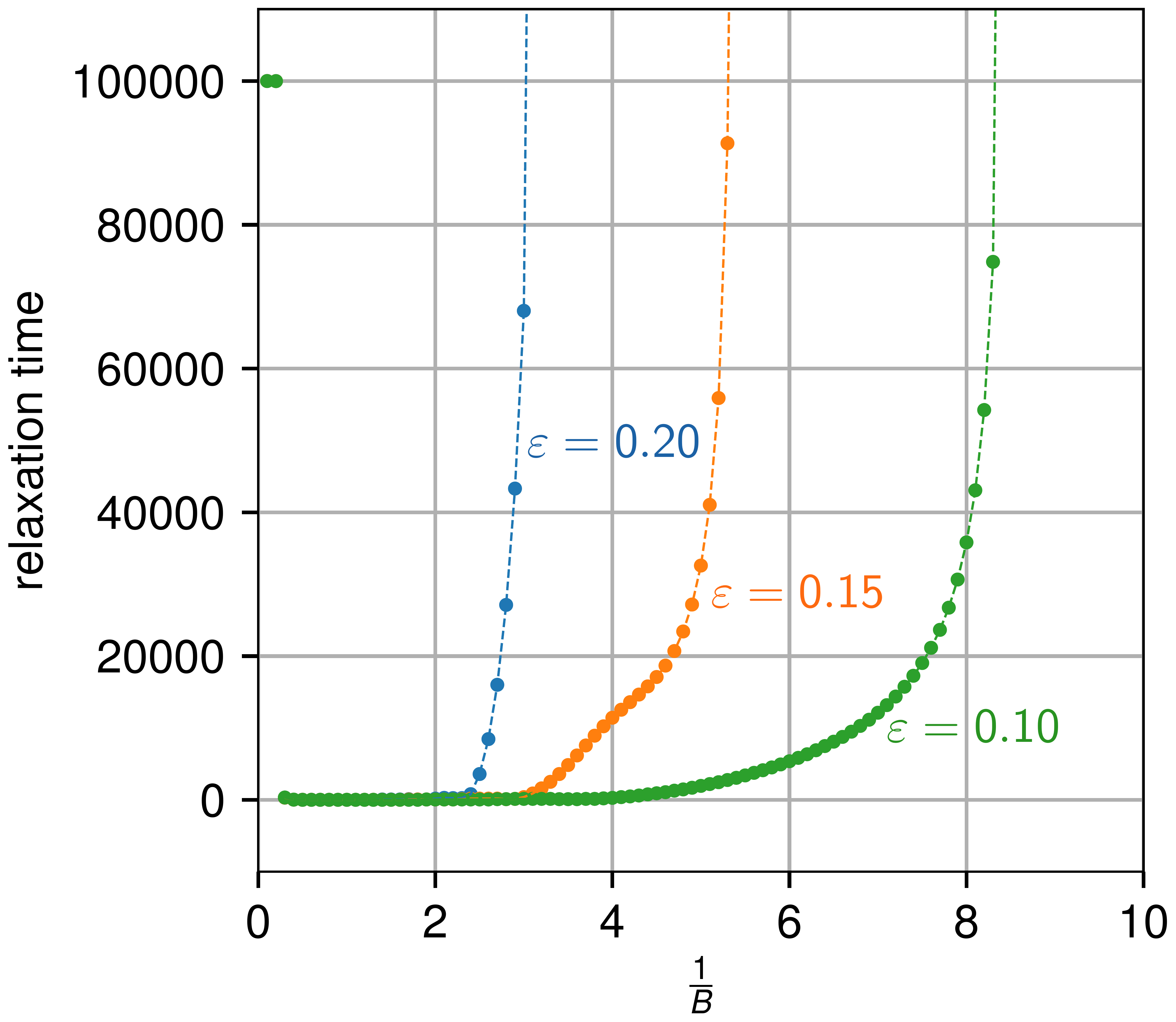}
\caption{
Relaxation time $\tau$ as function of $1/B$ as in Fig.\ \ref{fig:met} but for an insulator, see Eq.\ (\ref{eq:hamins}), and for different values of the on-site potential $\varepsilon$ as indicated.
} 
\label{fig:ins}
\end{figure}

Thus, in the present context, Eq.\ (\ref{eq:taullg}) is not expected to capture the case of very strong fields. 
For strong $B$, the field term will eventually dominate and only the precessional motion will survive. 
This means that $\tau$ should increase with increasing $B$ and diverge for $B\to \infty$.
In fact, as is seen in Fig.\ \ref{fig:met} for field strengths exceeding a critical strength of the order of the band width, the computed relaxation time diverges.


On the other hand, Eq.\ (\ref{eq:taullg}) should well describe the physics at weak $B$. 
It is satisfying to note that our approach, based on microscopic calculations including the details of the electronic structure perfectly agrees with the prediction of the spin-only LLG theory. 
As is seen in the figure, the relaxation time is proportional to $1/B$ for weak fields down to $B=0.0001$.
We conclude that even for very moderate system sizes $L$ and even for times scales of the order of $10^{5}$ inverse hoppings, the absorbing BC do not lead to any observable artifacts.

The predictive power can be exploited to study spin relaxation in cases where lowest-order perturbation theory in $J$ and the Markov approximation do not apply. 
One important example to be discussed here, is the case of a system with a gapped electronic structure. 
Even for a conventional band insulator, perturbation theory must break down, as this predicts the Gilbert damping constant to be given by \cite{BNF12,SH03,SP15} 
\be
\alpha = J^{2} \frac{\partial}{\partial \omega} \mbox{Im} \chi^{\rm (ret)}(\omega) \Big|_{\omega=0} \: . 
\ee
For an insulator with a gapped electronic structure, the imaginary part of the retarded magnetic susceptibility $\chi^{\rm (ret)}(\omega)$ must vanish in a finite range of excitation energies $\omega$ around $\omega=0$, which immediately implies $\alpha = 0$.
Hence, perturbation theory predicts the absence of damping, i.e., an infinite spin-relaxation time, independent of the field strength. 
However, this is unphysical since relaxation should be possible, if the initially induced Larmor precession with frequency $\omega \approx B$ can couple to the magnetic modes in the electron system. 
This is the case when $\mbox{Im}\, \chi^{\rm (ret)}(\omega=B) \ne 0$, i.e., for field strengths of the order of the fundamental gap or larger. 
Hence, a more elaborate effective theory would be necessary to cover this case.

The microscopic theory that includes the electronic degrees of freedom explicitly, on the other hand, perfectly complies with the expectation of a critical field strength:
Fig.\ \ref{fig:ins} displays results for the spin-relaxation time $\tau$ as obtained for a simple one-dimensional model of a band insulator, which is constructed by replacing 
\be
T_{ii'} \mapsto T_{ii'} + \varepsilon_{0} (-1)^{i} \delta_{ii'}
\label{eq:hamins}
\ee
in the Hamiltonian, Eq.\ (\ref{eq:ham}), or, equivalently, in the effective hopping matrix, Eq.\ (\ref{eq:teff}). 
The staggered on-site potential of strength $\varepsilon_{0}>0$ leads to a doubling of the unit cell and opens a gap of size $\Delta E = 2 \varepsilon_{0}$ in the bulk band structure at the edges of the reduced Brillouin zone. 
Here, for a finite system, the gap is $\Delta E \gtrsim 2 \varepsilon_{0}$. 
For $L=46$ sites, however, the difference is small, and we have checked that the results do not change significantly when increasing $L$.
Fig.\ \ref{fig:ins} indeed shows that complete spin relaxation is possible if the spin is driven with a sufficiently strong field. 
A divergent spin-relaxation time ($\tau > 100,000$) is only found for field strengths weaker than a certain critical value related to the gap size.

\begin{figure}[t]
\centering
\includegraphics[width=0.8\columnwidth]{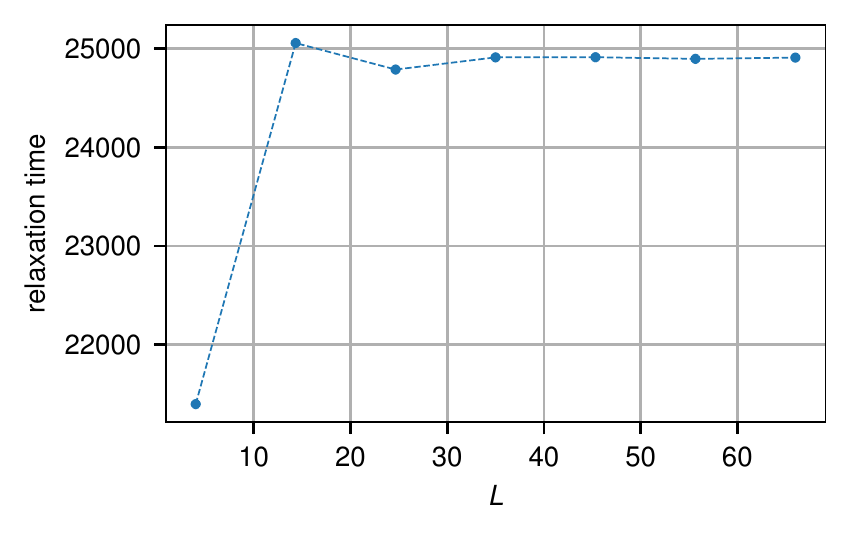}
\caption{
Relaxation time $\tau$ as function of the system size $L = 6, 16, 26, 36, 46, 56, 66$ and  $L_{B}=5=\mbox{const}$ as in Fig.\ \ref{fig:met} for $B=0.001$.
} 
\label{fig:ldep}
\end{figure}

Finally, we would like to stress that our approach is a systematic one, as the reliability of the approximations involved is fully controlled by the choice for the size of the system $L$. 
For $L\to \infty$, one trivially recovers the exact dynamics of a spin and of the coupled electron degrees of freedom, since the Lindblad-type boundaries becomes meaningless and since the construction of the boundaries is the only approximative element of the theory. 
Thus, varying the system size gives a good impression on the quality of results. 

To give an example, we display in Fig.\ \ref{fig:ldep} the spin relaxation time $\tau$ at a fixed field strength $B=0.001$ but as a function of $L$ for the metallic case. 
This corresponds to Fig.\ \ref{fig:met} where $L=46$ was chosen to represent converged results. 
Fig.\ \ref{fig:ldep} demonstrates that this is in fact the case:
We have $\tau \approx 25,000$ for $L=46$, and this value is not significantly changing when larger system sizes are considered.
For $L=56$ and $L=66$, we get the same value from the numerical calculation within an error of less than $0.1\%$.
It is very satisfying to see that already $L=16$ sites are actually quite sufficient, and only with $L=6$, which means one site that is left unchanged plus 5 sites coupling to the bath, the deviation of about $15\%$ is clearly beyond what should be tolerated.

\section{Conclusions}
\label{sec:con}

The real-time dynamics of local magnetic moments interacting with a large conduction-electron system is in most cases much slower than the bare electronic time scales. 
One general reason is the weakness of magnetic interactions compared to the conduction-band width or to the Fermi energy. 
Moreover, spin dynamics can be slowed down by missing phase space for magnetic scattering or by strongly anisotropic magnetic interactions and by other effects.
The strong separation of energy and time scales makes the theoretical description a challenging task.
For the study of relaxation phenomena, for example, it is the long-time limit that is of primary interest, but this cannot be treated independently from and is actually governed by the fast electronic processes.

On short time scales, perturbation theory, exploiting the separation of energy scales, can be very helpful. 
Master-equation approaches, including the Landau-Lifschitz-Gilbert approach, Redfield and other more sophisticated theories are quite powerful but are necessarily based on approximations, which in most cases are of {\em ad hoc} character and can be controlled {\em a posteriori} only.
For complex dynamics with phase-space bottlenecks, pre-relaxation phenomena or emergent symmetries, there is clearly an urgent need for a fully atomistic modelling, which covers time scales spanning several orders of magnitude and which is controlled {\em systematically}.

Here, we have presented the first steps towards such an approach.
The main idea is that relaxation processes are unidirectional, i.e., they are characterized by dissipation of energy and other conserved quantities due to flows of energy, spin etc.\ away from the initially excited core system to the electronic bulk but not vice versa. 
The fast processes in the core system, consisting of the local magnetic moments and the immediate surrounding, lead to the emission of wave packets carrying energy and spin, which implies that the core region must approach its ground state in the course of time.
Hence, the theory must (i) treat this spatial region exactly and (ii) must ensure that the processes within the core region and the excitations leaving the core region are not disturbed by artificial excitations back-propagating from the bulk to the core. 
Those back-propagating modes, however, are typically produced by reflections of outgoing wave packets from the edges of a system of finite extent, i.e., they result from the use of open or periodic boundary conditions. 

Boundary conditions, which fully absorb the outgoing excitations, solve the problem. 
We found that those can be realized with coupling the sites that are close to the edges of the finite system to an external bath as described by the Lindblad equation for the one-particle reduced density matrix. 
The important point is that the master-equation approach is merely employed as a technical tool to realize the absorbing boundaries while the quality of the approximation is solely controlled by the size of the core region, and, thus, we get a systematic approach. 

It has turned out that the Lindblad coupling to the bath does a perfect job inasmuch as the absorption is concerned. 
However, the naive implemention of Lindblad boundaries also {\em generates} excitations propagating from the edges to the core right at the start of the dynamics. 
Fortunately, this problem could be solved completely by using a Lindblad approach with matrix-valued Lindblad parameters that are fixed to perfectly suppress the mentioned initial-state artifacts.

This novel type of absorbing boundaries has been tested in detail. 
For a single classical spin coupled to a one-dimensional system of conduction electrons we were easily able to trace the atomistic real-time dynamics on a time scale longer than $10^{5}$ inverse hoppings without any noticeable problem. 
The computational limitation is solely given by the necessary size $L$ of the core system. 
For the currently studied case, we find that $L \lesssim 50$ is fully sufficient for convergence of the results. 

Future applications will address systems with several spins, coupled to electron systems in two and three dimensions, and including anisotropic interactions. 
The role of lattice degrees of freedom could be investigated as well. 
For quantitative and realistic studies, relaxation mediated also via phonons is an important aspect. 
Work along these lines is in progress. 
An open question is whether correlated electron systems might be treated within a similar framework on a level beyond standard Hartree-Fock theory. 
\\

\acknowledgments
This work was supported by the Deutsche Forschungsgemeinschaft (DFG) through the Cluster of Excellence ``Advanced Imaging of Matter'' - EXC 2056 - project ID 390715994, and by the DFG 
Sonderforschungsbereich 925 ``Light-induced dynamics and control of correlated quantum systems''
(project B5).

%


\begin{thebibliography}{38}%
\makeatletter
\providecommand \@ifxundefined [1]{%
 \@ifx{#1\undefined}
}%
\providecommand \@ifnum [1]{%
 \ifnum #1\expandafter \@firstoftwo
 \else \expandafter \@secondoftwo
 \fi
}%
\providecommand \@ifx [1]{%
 \ifx #1\expandafter \@firstoftwo
 \else \expandafter \@secondoftwo
 \fi
}%
\providecommand \natexlab [1]{#1}%
\providecommand \enquote  [1]{``#1''}%
\providecommand \bibnamefont  [1]{#1}%
\providecommand \bibfnamefont [1]{#1}%
\providecommand \citenamefont [1]{#1}%
\providecommand \href@noop [0]{\@secondoftwo}%
\providecommand \href [0]{\begingroup \@sanitize@url \@href}%
\providecommand \@href[1]{\@@startlink{#1}\@@href}%
\providecommand \@@href[1]{\endgroup#1\@@endlink}%
\providecommand \@sanitize@url [0]{\catcode `\\12\catcode `\$12\catcode
  `\&12\catcode `\#12\catcode `\^12\catcode `\_12\catcode `\%12\relax}%
\providecommand \@@startlink[1]{}%
\providecommand \@@endlink[0]{}%
\providecommand \url  [0]{\begingroup\@sanitize@url \@url }%
\providecommand \@url [1]{\endgroup\@href {#1}{\urlprefix }}%
\providecommand \urlprefix  [0]{URL }%
\providecommand \Eprint [0]{\href }%
\providecommand \doibase [0]{https://doi.org/}%
\providecommand \selectlanguage [0]{\@gobble}%
\providecommand \bibinfo  [0]{\@secondoftwo}%
\providecommand \bibfield  [0]{\@secondoftwo}%
\providecommand \translation [1]{[#1]}%
\providecommand \BibitemOpen [0]{}%
\providecommand \bibitemStop [0]{}%
\providecommand \bibitemNoStop [0]{.\EOS\space}%
\providecommand \EOS [0]{\spacefactor3000\relax}%
\providecommand \BibitemShut  [1]{\csname bibitem#1\endcsname}%
\let\auto@bib@innerbib\@empty
\bibitem [{\citenamefont {Tatara}\ \emph {et~al.}(2008)\citenamefont {Tatara},
  \citenamefont {Kohno},\ and\ \citenamefont {Shibata}}]{TKS08}%
  \BibitemOpen
  \bibfield  {author} {\bibinfo {author} {\bibfnamefont {G.}~\bibnamefont
  {Tatara}}, \bibinfo {author} {\bibfnamefont {H.}~\bibnamefont {Kohno}},\ and\
  \bibinfo {author} {\bibfnamefont {J.}~\bibnamefont {Shibata}},\ }\href@noop
  {} {\bibfield  {journal} {\bibinfo  {journal} {Physics Reports}\ }\textbf
  {\bibinfo {volume} {468}},\ \bibinfo {pages} {213} (\bibinfo {year}
  {2008})}\BibitemShut {NoStop}%
\bibitem [{\citenamefont {Skubic}\ \emph {et~al.}(2008)\citenamefont {Skubic},
  \citenamefont {Hellsvik}, \citenamefont {Nordstr\"om},\ and\ \citenamefont
  {Eriksson}}]{SHNE08}%
  \BibitemOpen
  \bibfield  {author} {\bibinfo {author} {\bibfnamefont {B.}~\bibnamefont
  {Skubic}}, \bibinfo {author} {\bibfnamefont {J.}~\bibnamefont {Hellsvik}},
  \bibinfo {author} {\bibfnamefont {L.}~\bibnamefont {Nordstr\"om}},\ and\
  \bibinfo {author} {\bibfnamefont {O.}~\bibnamefont {Eriksson}},\ }\href@noop
  {} {\bibfield  {journal} {\bibinfo  {journal} {J. Phys.: Condens. Matter}\
  }\textbf {\bibinfo {volume} {20}},\ \bibinfo {pages} {315203} (\bibinfo
  {year} {2008})}\BibitemShut {NoStop}%
\bibitem [{\citenamefont {Bertotti}\ \emph {et~al.}(2009)\citenamefont
  {Bertotti}, \citenamefont {Mayergoyz},\ and\ \citenamefont
  {Serpico}}]{BMS09}%
  \BibitemOpen
  \bibfield  {author} {\bibinfo {author} {\bibfnamefont {G.}~\bibnamefont
  {Bertotti}}, \bibinfo {author} {\bibfnamefont {I.~D.}\ \bibnamefont
  {Mayergoyz}},\ and\ \bibinfo {author} {\bibfnamefont {C.}~\bibnamefont
  {Serpico}}},
  {\em Nonlinear Magnetization Dynamics in Nanosystems}
  (\bibinfo  {publisher} {Elsevier},\ \bibinfo
  {address} {Amsterdam},\ \bibinfo {year} {2009})\BibitemShut {NoStop}%
\bibitem [{\citenamefont {F\"ahnle}\ and\ \citenamefont {Illg}(2011)}]{FI11}%
  \BibitemOpen
  \bibfield  {author} {\bibinfo {author} {\bibfnamefont {M.}~\bibnamefont
  {F\"ahnle}}\ and\ \bibinfo {author} {\bibfnamefont {C.}~\bibnamefont
  {Illg}},\ }\href@noop {} {\bibfield  {journal} {\bibinfo  {journal} {J.
  Phys.: Condens. Matter}\ }\textbf {\bibinfo {volume} {23}},\ \bibinfo {pages}
  {493201} (\bibinfo {year} {2011})}\BibitemShut {NoStop}%
\bibitem [{\citenamefont {Evans}\ \emph {et~al.}(2014)\citenamefont {Evans},
  \citenamefont {Fan}, \citenamefont {Chureemart}, \citenamefont {Ostler},
  \citenamefont {Ellis},\ and\ \citenamefont {Chantrell}}]{EFC+14}%
  \BibitemOpen
  \bibfield  {author} {\bibinfo {author} {\bibfnamefont {R.~F.~L.}\
  \bibnamefont {Evans}}, \bibinfo {author} {\bibfnamefont {W.~J.}\ \bibnamefont
  {Fan}}, \bibinfo {author} {\bibfnamefont {P.}~\bibnamefont {Chureemart}},
  \bibinfo {author} {\bibfnamefont {T.~A.}\ \bibnamefont {Ostler}}, \bibinfo
  {author} {\bibfnamefont {M.~O.~A.}\ \bibnamefont {Ellis}},\ and\ \bibinfo
  {author} {\bibfnamefont {R.~W.}\ \bibnamefont {Chantrell}},\ }\href@noop {}
  {\bibfield  {journal} {\bibinfo  {journal} {J. Phys.: Condens. Matter}\
  }\textbf {\bibinfo {volume} {26}},\ \bibinfo {pages} {103202} (\bibinfo
  {year} {2014})}\BibitemShut {NoStop}%
\bibitem [{llg()}]{llg}%
  \BibitemOpen
  \href@noop {} {}\bibinfo {note} {L.~D. Landau and E.~M. Lifshitz, Physik.
  Zeits. Sowjetunion \textbf{8},153 (1935); T. Gilbert, Phys. Rev.
  \textbf{100}, 1243 (1955); T. Gilbert, Magnetics, IEEE Transactions on
  \textbf{40}, 3443 (2004).}\BibitemShut {Stop}%
\bibitem [{VZ()}]{VZ}%
  \BibitemOpen
  \href@noop {} {}\bibinfo {note} {S. V. Vonsovsky, Zh. \'Eksp. Teor. Fiz.
  {\bfseries 16}, 981 (1946); C. Zener, Phys. Rev. {\bfseries 81}, 440 (1951);
  S. V. Vonsovsky and E. A. Turov, Zh. \'Eksp. Teor. Fiz. {\bfseries 24}, 419
  (1953).}\BibitemShut {Stop}%
\bibitem [{\citenamefont {Onoda}\ and\ \citenamefont {Nagaosa}(2006)}]{ON06}%
  \BibitemOpen
  \bibfield  {author} {\bibinfo {author} {\bibfnamefont {M.}~\bibnamefont
  {Onoda}}\ and\ \bibinfo {author} {\bibfnamefont {N.}~\bibnamefont
  {Nagaosa}},\ }\href@noop {} {\bibfield  {journal} {\bibinfo  {journal} {Phys.
  Rev. Lett.}\ }\textbf {\bibinfo {volume} {96}},\ \bibinfo {pages} {066603}
  (\bibinfo {year} {2006})}\BibitemShut {NoStop}%
\bibitem [{\citenamefont {Bhattacharjee}\ \emph {et~al.}(2012)\citenamefont
  {Bhattacharjee}, \citenamefont {Nordstr\"om},\ and\ \citenamefont
  {Fransson}}]{BNF12}%
  \BibitemOpen
  \bibfield  {author} {\bibinfo {author} {\bibfnamefont {S.}~\bibnamefont
  {Bhattacharjee}}, \bibinfo {author} {\bibfnamefont {L.}~\bibnamefont
  {Nordstr\"om}},\ and\ \bibinfo {author} {\bibfnamefont {J.}~\bibnamefont
  {Fransson}},\ }\href@noop {} {\bibfield  {journal} {\bibinfo  {journal}
  {Phys. Rev. Lett.}\ }\textbf {\bibinfo {volume} {108}},\ \bibinfo {pages}
  {057204} (\bibinfo {year} {2012})}\BibitemShut {NoStop}%
\bibitem [{\citenamefont {Umetsu}\ \emph {et~al.}(2012)\citenamefont {Umetsu},
  \citenamefont {Miura},\ and\ \citenamefont {Sakuma}}]{UMS12}%
  \BibitemOpen
  \bibfield  {author} {\bibinfo {author} {\bibfnamefont {N.}~\bibnamefont
  {Umetsu}}, \bibinfo {author} {\bibfnamefont {D.}~\bibnamefont {Miura}},\ and\
  \bibinfo {author} {\bibfnamefont {A.}~\bibnamefont {Sakuma}},\ }\href@noop {}
  {\bibfield  {journal} {\bibinfo  {journal} {J. Appl. Phys.}\ }\textbf
  {\bibinfo {volume} {111}},\ \bibinfo {eid} {07D117} (\bibinfo {year}
  {2012})}\BibitemShut {NoStop}%
\bibitem [{\citenamefont {Bajpai}\ and\ \citenamefont {Nikolic}(2019)}]{BN19}%
  \BibitemOpen
  \bibfield  {author} {\bibinfo {author} {\bibfnamefont {U.}~\bibnamefont
  {Bajpai}}\ and\ \bibinfo {author} {\bibfnamefont {B.~K.}\ \bibnamefont
  {Nikolic}},\ }\href@noop {} {\bibfield  {journal} {\bibinfo  {journal} {Phys.
  Rev. B}\ }\textbf {\bibinfo {volume} {99}},\ \bibinfo {pages} {134409}
  (\bibinfo {year} {2019})}\BibitemShut {NoStop}%
\bibitem [{\citenamefont {Antropov}\ \emph {et~al.}(1995)\citenamefont
  {Antropov}, \citenamefont {Katsnelson}, \citenamefont {van Schilfgaarde},\
  and\ \citenamefont {Harmon}}]{AKvSH95}%
  \BibitemOpen
  \bibfield  {author} {\bibinfo {author} {\bibfnamefont {V.~P.}\ \bibnamefont
  {Antropov}}, \bibinfo {author} {\bibfnamefont {M.~I.}\ \bibnamefont
  {Katsnelson}}, \bibinfo {author} {\bibfnamefont {M.}~\bibnamefont {van
  Schilfgaarde}},\ and\ \bibinfo {author} {\bibfnamefont {B.~N.}\ \bibnamefont
  {Harmon}},\ }\href@noop {} {\bibfield  {journal} {\bibinfo  {journal} {Phys.
  Rev. Lett.}\ }\textbf {\bibinfo {volume} {75}},\ \bibinfo {pages} {729}
  (\bibinfo {year} {1995})}\BibitemShut {NoStop}%
\bibitem [{\citenamefont {Kune\v{s}}\ and\ \citenamefont
  {Kambersk\'y}(2002)}]{KK02}%
  \BibitemOpen
  \bibfield  {author} {\bibinfo {author} {\bibfnamefont {J.}~\bibnamefont
  {Kune\v{s}}}\ and\ \bibinfo {author} {\bibfnamefont {V.}~\bibnamefont
  {Kambersk\'y}},\ }\href@noop {} {\bibfield  {journal} {\bibinfo  {journal}
  {Phys. Rev. B}\ }\textbf {\bibinfo {volume} {65}},\ \bibinfo {pages} {212411}
  (\bibinfo {year} {2002})}\BibitemShut {NoStop}%
\bibitem [{\citenamefont {Capelle}\ and\ \citenamefont {Gyorffy}(2003)}]{CG03}%
  \BibitemOpen
  \bibfield  {author} {\bibinfo {author} {\bibfnamefont {K.}~\bibnamefont
  {Capelle}}\ and\ \bibinfo {author} {\bibfnamefont {B.~L.}\ \bibnamefont
  {Gyorffy}},\ }\href@noop {} {\bibfield  {journal} {\bibinfo  {journal}
  {Europhys. Lett.}\ }\textbf {\bibinfo {volume} {61}},\ \bibinfo {pages} {354}
  (\bibinfo {year} {2003})}\BibitemShut {NoStop}%
\bibitem [{\citenamefont {Ebert}\ \emph {et~al.}(2011)\citenamefont {Ebert},
  \citenamefont {Mankovsky}, \citenamefont {K\"odderitzsch},\ and\
  \citenamefont {Kelly}}]{EMKK11}%
  \BibitemOpen
  \bibfield  {author} {\bibinfo {author} {\bibfnamefont {H.}~\bibnamefont
  {Ebert}}, \bibinfo {author} {\bibfnamefont {S.}~\bibnamefont {Mankovsky}},
  \bibinfo {author} {\bibfnamefont {D.}~\bibnamefont {K\"odderitzsch}},\ and\
  \bibinfo {author} {\bibfnamefont {P.~J.}\ \bibnamefont {Kelly}},\ }\href@noop
  {} {\bibfield  {journal} {\bibinfo  {journal} {Phys. Rev. Lett.}\ }\textbf
  {\bibinfo {volume} {107}},\ \bibinfo {pages} {066603} (\bibinfo {year}
  {2011})}\BibitemShut {NoStop}%
\bibitem [{\citenamefont {Sakuma}(2012)}]{Sak12}%
  \BibitemOpen
  \bibfield  {author} {\bibinfo {author} {\bibfnamefont {A.}~\bibnamefont
  {Sakuma}},\ }\href@noop {} {\bibfield  {journal} {\bibinfo  {journal} {J.
  Phys. Soc. Jpn.}\ }\textbf {\bibinfo {volume} {81}},\ \bibinfo {pages}
  {084701} (\bibinfo {year} {2012})}\BibitemShut {NoStop}%
\bibitem [{\citenamefont {Sayad}\ and\ \citenamefont {Potthoff}(2015)}]{SP15}%
  \BibitemOpen
  \bibfield  {author} {\bibinfo {author} {\bibfnamefont {M.}~\bibnamefont
  {Sayad}}\ and\ \bibinfo {author} {\bibfnamefont {M.}~\bibnamefont
  {Potthoff}},\ }\href@noop {} {\bibfield  {journal} {\bibinfo  {journal} {New
  J. Phys.}\ }\textbf {\bibinfo {volume} {17}},\ \bibinfo {pages} {113058}
  (\bibinfo {year} {2015})}\BibitemShut {NoStop}%
\bibitem [{\citenamefont {Sayad}\ \emph {et~al.}(2016)\citenamefont {Sayad},
  \citenamefont {Rausch},\ and\ \citenamefont {Potthoff}}]{SRP16a}%
  \BibitemOpen
  \bibfield  {author} {\bibinfo {author} {\bibfnamefont {M.}~\bibnamefont
  {Sayad}}, \bibinfo {author} {\bibfnamefont {R.}~\bibnamefont {Rausch}},\ and\
  \bibinfo {author} {\bibfnamefont {M.}~\bibnamefont {Potthoff}},\ }\href@noop
  {} {\bibfield  {journal} {\bibinfo  {journal} {Phys. Rev. Lett.}\ }\textbf
  {\bibinfo {volume} {117}},\ \bibinfo {pages} {127201} (\bibinfo {year}
  {2016})}\BibitemShut {NoStop}%
\bibitem [{\citenamefont {Stahl}\ and\ \citenamefont {Potthoff}(2017)}]{SP17}%
  \BibitemOpen
  \bibfield  {author} {\bibinfo {author} {\bibfnamefont {C.}~\bibnamefont
  {Stahl}}\ and\ \bibinfo {author} {\bibfnamefont {M.}~\bibnamefont
  {Potthoff}},\ }\href@noop {} {\bibfield  {journal} {\bibinfo  {journal}
  {Phys. Rev. Lett.}\ }\textbf {\bibinfo {volume} {119}},\ \bibinfo {pages}
  {227203} (\bibinfo {year} {2017})}\BibitemShut {NoStop}%
\bibitem [{\citenamefont {Elbracht}\ \emph {et~al.}(2020)\citenamefont
  {Elbracht}, \citenamefont {Michel},\ and\ \citenamefont {Potthoff}}]{EMP20}%
  \BibitemOpen
  \bibfield  {author} {\bibinfo {author} {\bibfnamefont {M.}~\bibnamefont
  {Elbracht}}, \bibinfo {author} {\bibfnamefont {S.}~\bibnamefont {Michel}},\
  and\ \bibinfo {author} {\bibfnamefont {M.}~\bibnamefont {Potthoff}},\
  }\href@noop {} {\bibfield  {journal} {\bibinfo  {journal} {Phys. Rev. Lett.}\
  }\textbf {\bibinfo {volume} {124}},\ \bibinfo {pages} {197202} (\bibinfo
  {year} {2020})}\BibitemShut {NoStop}%
\bibitem [{\citenamefont {Bajpai}\ and\ \citenamefont {Nikolic}(2020)}]{BN20}%
  \BibitemOpen
  \bibfield  {author} {\bibinfo {author} {\bibfnamefont {U.}~\bibnamefont
  {Bajpai}}\ and\ \bibinfo {author} {\bibfnamefont {B.~K.}\ \bibnamefont
  {Nikolic}},\ }\href@noop {} {arXiv:2005.14153}
  \BibitemShut {NoStop}%
\bibitem [{\citenamefont {Antoine}\ \emph {et~al.}(2008)\citenamefont
  {Antoine}, \citenamefont {Arnold}, \citenamefont {Besse}, \citenamefont
  {Ehrhardt},\ and\ \citenamefont {Schadle}}]{AAB+08}%
  \BibitemOpen
  \bibfield  {author} {\bibinfo {author} {\bibfnamefont {X.}~\bibnamefont
  {Antoine}}, \bibinfo {author} {\bibfnamefont {A.}~\bibnamefont {Arnold}},
  \bibinfo {author} {\bibfnamefont {C.}~\bibnamefont {Besse}}, \bibinfo
  {author} {\bibfnamefont {M.}~\bibnamefont {Ehrhardt}},\ and\ \bibinfo
  {author} {\bibfnamefont {A.}~\bibnamefont {Schadle}},\ }\href@noop {}
  {\bibfield  {journal} {\bibinfo  {journal} {Commun. Comput. Phys.}\ }\textbf
  {\bibinfo {volume} {4}},\ \bibinfo {pages} {729} (\bibinfo {year}
  {2008})}\BibitemShut {NoStop}%
\bibitem [{\citenamefont {Manolopoulos}(2002)}]{Man02}%
  \BibitemOpen
  \bibfield  {author} {\bibinfo {author} {\bibfnamefont {D.~E.}\ \bibnamefont
  {Manolopoulos}},\ }\href@noop {} {\bibfield  {journal} {\bibinfo  {journal}
  {J. Chem. Phys.}\ }\textbf {\bibinfo {volume} {117}},\ \bibinfo {pages}
  {9552} (\bibinfo {year} {2002})}\BibitemShut {NoStop}%
\bibitem [{\citenamefont {Berenger}(1994)}]{Ber94}%
  \BibitemOpen
  \bibfield  {author} {\bibinfo {author} {\bibfnamefont {J.}~\bibnamefont
  {Berenger}},\ }\href@noop {} {\bibfield  {journal} {\bibinfo  {journal} {J.
  Comput. Phys.}\ }\textbf {\bibinfo {volume} {114}},\ \bibinfo {pages} {185}
  (\bibinfo {year} {1994})}\BibitemShut {NoStop}%
\bibitem [{\citenamefont {Selst\o}\ and\ \citenamefont {Kvaal}(2010)}]{SK10}%
  \BibitemOpen
  \bibfield  {author} {\bibinfo {author} {\bibfnamefont {S.}~\bibnamefont
  {Selst\o}}\ and\ \bibinfo {author} {\bibfnamefont {S.}~\bibnamefont
  {Kvaal}},\ }\href@noop {} {\bibfield  {journal} {\bibinfo  {journal} {J.
  Phys. B}\ }\textbf {\bibinfo {volume} {43}},\ \bibinfo {pages} {065004}
  (\bibinfo {year} {2010})}\BibitemShut {NoStop}%
\bibitem [{\citenamefont {Lindblad}(1976)}]{Lin76}%
  \BibitemOpen
  \bibfield  {author} {\bibinfo {author} {\bibfnamefont {G.}~\bibnamefont
  {Lindblad}},\ }\href@noop {} {\bibfield  {journal} {\bibinfo  {journal}
  {Commun. Math. Phys.}\ }\textbf {\bibinfo {volume} {48}},\ \bibinfo {pages}
  {119} (\bibinfo {year} {1976})}\BibitemShut {NoStop}%
\bibitem [{\citenamefont {Pearle}(2012)}]{Pea12}%
  \BibitemOpen
  \bibfield  {author} {\bibinfo {author} {\bibfnamefont {P.}~\bibnamefont
  {Pearle}},\ }\href@noop {} {\bibfield  {journal} {\bibinfo  {journal} {Eur.
  J. Phys.}\ }\textbf {\bibinfo {volume} {33}},\ \bibinfo {pages} {805}
  (\bibinfo {year} {2012})}\BibitemShut {NoStop}%
\bibitem [{\citenamefont {Carmichael}(1993)}]{Car93}%
  \BibitemOpen
  \bibfield  {author} {\bibinfo {author} {\bibfnamefont {H.}~\bibnamefont
  {Carmichael}},\ }\href@noop {} {\emph {\bibinfo {title} {An Open Systems
  Approach to Quantum Optics}}}\ (\bibinfo  {publisher} {Springer},\ \bibinfo
  {address} {Berlin},\ \bibinfo {year} {1993})\BibitemShut {NoStop}%
\bibitem [{\citenamefont {Breuer}\ and\ \citenamefont
  {Petruccione}(2010)}]{BP10b}%
  \BibitemOpen
  \bibfield  {author} {\bibinfo {author} {\bibfnamefont {H.-P.}\ \bibnamefont
  {Breuer}}\ and\ \bibinfo {author} {\bibfnamefont {F.}~\bibnamefont
  {Petruccione}},\ }\href@noop {} {\emph {\bibinfo {title} {The Theory of Open
  Quantum Systems}}}\ (\bibinfo  {publisher} {Oxford Univ. Press},\ \bibinfo
  {address} {Oxford},\ \bibinfo {year} {2010})\BibitemShut {NoStop}%
\bibitem [{\citenamefont {Xu}\ \emph {et~al.}(2019)\citenamefont {Xu},
  \citenamefont {Thingna}, \citenamefont {Guo},\ and\ \citenamefont
  {Poletti}}]{XTGP19}%
  \BibitemOpen
  \bibfield  {author} {\bibinfo {author} {\bibfnamefont {X.}~\bibnamefont
  {Xu}}, \bibinfo {author} {\bibfnamefont {J.}~\bibnamefont {Thingna}},
  \bibinfo {author} {\bibfnamefont {C.}~\bibnamefont {Guo}},\ and\ \bibinfo
  {author} {\bibfnamefont {D.}~\bibnamefont {Poletti}},\ }\href@noop {}
  {\bibfield  {journal} {\bibinfo  {journal} {Phys. Rev. A}\ }\textbf {\bibinfo
  {volume} {99}},\ \bibinfo {pages} {012106} (\bibinfo {year}
  {2019})}\BibitemShut {NoStop}%
\bibitem [{\citenamefont {Arrigoni}\ \emph {et~al.}(2013)\citenamefont
  {Arrigoni}, \citenamefont {Knap},\ and\ \citenamefont {von~der
  Linden}}]{AKvdL13}%
  \BibitemOpen
  \bibfield  {author} {\bibinfo {author} {\bibfnamefont {E.}~\bibnamefont
  {Arrigoni}}, \bibinfo {author} {\bibfnamefont {M.}~\bibnamefont {Knap}},\
  and\ \bibinfo {author} {\bibfnamefont {W.}~\bibnamefont {von~der Linden}},\
  }\href@noop {} {\bibfield  {journal} {\bibinfo  {journal} {Phys. Rev. Lett.}\
  }\textbf {\bibinfo {volume} {110}},\ \bibinfo {pages} {086403} (\bibinfo
  {year} {2013})}\BibitemShut {NoStop}%
\bibitem [{\citenamefont {Dzhioev}\ and\ \citenamefont {Kosov}(2011)}]{DK11}%
  \BibitemOpen
  \bibfield  {author} {\bibinfo {author} {\bibfnamefont {A.~A.}\ \bibnamefont
  {Dzhioev}}\ and\ \bibinfo {author} {\bibfnamefont {D.~S.}\ \bibnamefont
  {Kosov}},\ }\href@noop {} {\bibfield  {journal} {\bibinfo  {journal} {J.
  Chem. Phys.}\ }\textbf {\bibinfo {volume} {134}},\ \bibinfo {pages} {044121}
  (\bibinfo {year} {2011})}\BibitemShut {NoStop}%
\bibitem [{\citenamefont {Verstraete}\ \emph {et~al.}(2004)\citenamefont
  {Verstraete}, \citenamefont {Garc\'ia-Ripoll},\ and\ \citenamefont
  {Cirac}}]{VGRC04}%
  \BibitemOpen
  \bibfield  {author} {\bibinfo {author} {\bibfnamefont {F.}~\bibnamefont
  {Verstraete}}, \bibinfo {author} {\bibfnamefont {J.~J.}\ \bibnamefont
  {Garc\'ia-Ripoll}},\ and\ \bibinfo {author} {\bibfnamefont {J.~I.}\
  \bibnamefont {Cirac}},\ }\href@noop {} {\bibfield  {journal} {\bibinfo
  {journal} {Phys. Rev. Lett.}\ }\textbf {\bibinfo {volume} {93}},\ \bibinfo
  {pages} {207204} (\bibinfo {year} {2004})}\BibitemShut {NoStop}%
\bibitem [{\citenamefont {Zwolak}\ and\ \citenamefont {Vidal}(2004)}]{ZV04}%
  \BibitemOpen
  \bibfield  {author} {\bibinfo {author} {\bibfnamefont {M.}~\bibnamefont
  {Zwolak}}\ and\ \bibinfo {author} {\bibfnamefont {G.}~\bibnamefont {Vidal}},\
  }\href@noop {} {\bibfield  {journal} {\bibinfo  {journal} {Phys. Rev. Lett.}\
  }\textbf {\bibinfo {volume} {93}},\ \bibinfo {pages} {207205} (\bibinfo
  {year} {2004})}\BibitemShut {NoStop}%
\bibitem [{\citenamefont {Prosen}\ and\ \citenamefont {Znidari}(2009)}]{PZ09}%
  \BibitemOpen
  \bibfield  {author} {\bibinfo {author} {\bibfnamefont {T.}~\bibnamefont
  {Prosen}}\ and\ \bibinfo {author} {\bibfnamefont {M.}~\bibnamefont
  {Znidari}},\ }\href@noop {} {\bibfield  {journal} {\bibinfo  {journal} {J.
  Stat. Mech.},\ \bibinfo {pages} {P02035}} (\bibinfo {year}
  {2009})}\BibitemShut {NoStop}%
\bibitem [{\citenamefont {Elze}(2012)}]{Elz12}%
  \BibitemOpen
  \bibfield  {author} {\bibinfo {author} {\bibfnamefont {H.}~\bibnamefont
  {Elze}},\ }\href@noop {} {\bibfield  {journal} {\bibinfo  {journal} {Phys.
  Rev. A}\ }\textbf {\bibinfo {volume} {85}},\ \bibinfo {pages} {052109}
  (\bibinfo {year} {2012})}\BibitemShut {NoStop}%
\bibitem [{\citenamefont {Kikuchi}(1956)}]{Kik56}%
  \BibitemOpen
  \bibfield  {author} {\bibinfo {author} {\bibfnamefont {R.}~\bibnamefont
  {Kikuchi}},\ }\href@noop {} {\bibfield  {journal} {\bibinfo  {journal} {J.
  Appl. Phys.}\ }\textbf {\bibinfo {volume} {27}},\ \bibinfo {pages} {1352}
  (\bibinfo {year} {1956})}\BibitemShut {NoStop}%
\bibitem [{\citenamefont {Simanek}\ and\ \citenamefont
  {Heinrich}(2003)}]{SH03}%
  \BibitemOpen
  \bibfield  {author} {\bibinfo {author} {\bibfnamefont {E.}~\bibnamefont
  {Simanek}}\ and\ \bibinfo {author} {\bibfnamefont {B.}~\bibnamefont
  {Heinrich}},\ }\href@noop {} {\bibfield  {journal} {\bibinfo  {journal}
  {Phys. Rev. B}\ }\textbf {\bibinfo {volume} {67}},\ \bibinfo {pages} {144418}
  (\bibinfo {year} {2003})}\BibitemShut {NoStop}%
\end{thebibliography}

\end{document}